\documentclass[twocolumn]{aastex63}
\usepackage{graphicx}

\usepackage{hyperref}
\usepackage[colorinlistoftodos]{todonotes}
\usepackage{amsmath}
\usepackage{nicefrac}
\usepackage{bm}
\usepackage{comment}

\shorttitle{Thermal pressure deficit from tSZ}
\usepackage[encapsulated]{CJK}

\newcommand{\Mpch}{\ensuremath{\mathrm{Mpc}~h^{-1}}}

\newcommand{\Msol}{$\mathrm{M}_\odot$}

\newcommand{\eightsigma}{$f_\mathrm{gas}~-8\sigma$}

\newcommand{\CY}{Compton~$y$}
\newcommand{\fidsim}{L1\_m9}

\begin{document}

\title{Evidence for a thermal pressure deficit in galaxy groups from the tSZ effect and weak lensing}

\author[0000-0002-9337-0902]{Jared C. Siegel}
\altaffiliation{NSF Graduate Research Fellow}
\affiliation{Department of Astrophysical Sciences, Princeton University, 4 Ivy Lane, Princeton, NJ 08544, USA}
\email{siegeljc@princeton.edu}

\author[0000-0002-6445-0559]{Alexandra Amon}
\affiliation{Department of Astrophysical Sciences, Princeton University, 4 Ivy Lane, Princeton, NJ 08544, USA}

\author[0000-0002-5612-3427]{Jenny E. Greene}
\affiliation{Department of Astrophysical Sciences, Princeton University, 4 Ivy Lane, Princeton, NJ 08544, USA}

\author[0000-0002-1286-483X]{Ian G. McCarthy}
\affiliation{Astrophysics Research Institute, Liverpool John Moores University, Liverpool, L3 5RF, UK}

\author[0000-0001-9185-5044]{Eliot Quataert}
\affiliation{Department of Astrophysical Sciences, Princeton University, 4 Ivy Lane, Princeton, NJ 08544, USA}

\author[0000-0002-1297-3673]{William Coulton}
\affiliation{Kavli Institute for Cosmology Cambridge, Madingley Road, Cambridge CB3 0HA, UK}
\affiliation{DAMTP, Centre for Mathematical Sciences, University of Cambridge, Wilberforce Road, Cambridge CB3 0WA, UK}

\begin{abstract}
Measurements of the thermal Sunyaev-Zel'dovich (tSZ) effect have yet to form a consistent picture of the thermodynamic state of the gas in the intracluster medium:
their interpretation is complicated by foreground contamination and uncertain halo masses. We present new measurements of the tSZ effect around the Dark Energy Spectroscopic Instrument (DESI) Luminous Red Galaxy (LRG) sample, together with galaxy-galaxy lensing (GGL) measurements that enable a like-with-like comparison to state-of-the-art hydrodynamical simulations. 
We robustly isolate the tSZ signal by directly modeling the dust and radio emission of the target galaxies using the Atacama Cosmology Telescope (ACT) single-channel temperature maps, substantially reducing uncertainties from astrophysical foregrounds.
Across halo masses $M_{500}=10^{13}$--$10^{14}$~{\Msol} and redshifts $0.4<z<1$, we find that the fiducial $1$~Gpc$^3$ FLAMINGO simulation significantly overpredicts the observed tSZ signal at $\lesssim3'$ (i.e., $\lesssim 4\,R_{500}$ at $z=0.7$). Even the simulation with the strongest gas expulsion---which successfully reproduces the gas density inferred from kinetic SZ measurements of the same galaxy sample---overpredicts the thermal pressure. 
Because the strongest feedback model already reproduces the observed gas density, the remaining discrepancy is difficult to explain with additional gas depletion alone. 
Instead, current hydrodynamical simulations appear to overpredict the thermal pressure of galaxy groups by a factor of two, pointing toward missing non-thermal pressure support or significant departures from hydrostatic equilibrium.
\end{abstract}


\section{Introduction}
\label{sec:intro}

The nature of baryon feedback is imprinted in the intracluster
medium (ICM): the hot ionized gas in groups and clusters.
Supernovae, stellar winds, and active galactic nuclei (AGN)---the dominant feedback mechanism for massive galaxies, galaxy groups, and clusters \citep[e.g.,][]{McNamara2007,Conroy2008,McCarthy2011,Fabian2012,Gitti2012,Gaspari2020,Eckert2021}---heat and expel gas from halos.
Baryonic processes further alter the ICM through non-thermal pressure support, e.g., turbulent motions, magnetic fields,
cosmic rays, and plasma instabilities \citep[e.g.,][]{Rudd2009,Battaglia2012,Parrish2012,McCourt2013,Nelson2014,Quataert2025}.
The physics of these non-gravitational processes and their role in reshaping the ICM remain open questions.

Galaxy groups ($M_{500}=10^{13}$ to $10^{14}$~{\Msol}) are particularly sensitive laboratories for studying feedback: their shallower potential wells amplify the impact of non-gravitational processes compared to massive clusters \citep[e.g.,][]{McCarthy2010,Gaspari2019,Eckert2021,Eckert2024}.
Indeed, recent observations suggest that galaxy groups are substantially more gas depleted than previously thought \citep[e.g.,][]{Bigwood2024, Bulbul2024, Popesso2024, Hadzhiyska2024photoz, McCarthy2025, RiedGuachalla2025, Roper2025, Siegel2025flamingo, Bigwood2025allthesims}.
Evidence for efficient gas expulsion also comes from the suppression of the matter power spectrum relative to dark-matter-only predictions, now inferred from a range of observables
\citep{Schneider2022, Bigwood2024, Grandis2024, Ferreira2024, Kovac25, LaPosta2025, Pandey2025, Dalal2025, reischke2025, Siegel2025baryonification, Sharma2026}. 
Together, the growing landscape of feedback probes points to more dramatic baryon depletion from groups than is realized in most state-of-the-art hydrodynamical simulations.

The thermal state of the gas is less clear.
The thermal Sunyaev-Zel'dovich (tSZ) effect---sourced by the inverse Compton scattering of Cosmic Microwave Background (CMB) photons off the ionized gas around groups and clusters---traces the thermal pressure of the gas \citep[quantified by the {\CY} parameter;][]{Sunyaev1972}: $y \propto \int n_\mathrm{e} T_\mathrm{e}\,dl $.
From stacked measurements of the tSZ effect, previous studies have found agreement with the gravity-only expectation, pointing to a weak impact of feedback \citep{Planck2013tsz,Greco2015}; however, some recent studies have reported thermal pressure deficits relative to the gravity-only expectation \citep{Lim2018,Hill2018} and even some cosmological hydrodynamical simulations \citep{Das2023,Das2025,Liu2025}. 
Measurements of the tSZ power spectrum have also shown a deficit compared to simulation predictions \citep[e.g.,][]{McCarthy2014,Efstathiou2025,Raghunathan2026}.
Yet establishing a thermal pressure deficit from the tSZ effect is notoriously challenging from both a measurement and an interpretation perspective. 

In this work, we present new measurements of the tSZ signal around massive galaxies and compare with state-of-the-art hydrodynamical simulations \citep{Schaye2023}. 
We address the three principal roadblocks to probing baryon feedback with the tSZ effect:
(i) contamination of the tSZ signal by astrophysical foregrounds, e.g., dust emission and radio point sources; 
(ii) the sensitivity of the theory prediction to halo mass and satellite fraction;
and (iii) the degeneracy between density and temperature when interpreting the tSZ signal.

The tSZ signal is difficult to measure at group masses:
it is significantly fainter than the primary CMB anisotropies, and it is contaminated by emission from radio point sources and dust;
the cumulative dust emission from galaxies along the line of sight is known as the Cosmic Infrared Background (CIB). 
A common approach is to isolate the tSZ signal at the map level, constructing {\CY} maps from multi-frequency CMB data \citep[e.g.,][]{Bennett1992, Remazeilles2011, Hurier2013};
however, map-level separation must remove the emission of all sources along the line of sight without knowing their redshifts.
We therefore take a different approach, measuring the galaxies' stacked spectral energy distribution (SED) from the Atacama Cosmology Telescope (ACT) DR6 single-channel temperature maps \citep{Naess2020, Naess2025} and modeling the data with three components: tSZ, dust, and radio emission.
Background-subtracting the photometry suppresses uncorrelated foregrounds, so any contaminating emission must originate in the targeted galaxies themselves and/or their surrounding structure; critically, the galaxies' redshifts and radio flux limits are known.
This approach greatly simplifies the modeling challenge and leverages our knowledge of the targeted galaxies.
For comparison, we also present results from the dust ``deprojected'' ACT+Planck {\CY} maps of \cite{Coulton2024}.

The comparison of tSZ observations with theory is complicated by the particularly strong scaling of the tSZ effect with mass: in the self-similar (gravity-only) limit the tSZ signal scales as $M_\mathrm{halo}^{5/3}$ \citep{Kaiser1986}. 
Uncertain or biased halo mass estimates therefore masquerade as signs of feedback.
To address this challenge, we perform a like-with-like comparison between our new tSZ measurements and the $1$~Gpc$^{3}$ FLAMINGO hydrodynamical simulations \citep{Schaye2023, Kugel2023}, using new galaxy-galaxy lensing (GGL) measurements to calibrate the selection of simulated galaxies. 
The simulation comparison allows us to forward model the effects of satellites, miscentering, and two-halo contributions.

Even if the tSZ measurement is found to deviate from the model, the discrepancy could stem from differences in the gas density, temperature, or both, clouding insights into feedback. 
Because the kinetic Sunyaev-Zel'dovich (kSZ) effect probes the gas density \citep{Sunyaev1980}, this degeneracy can be broken for galaxy samples with existing kSZ measurements \citep[e.g.,][]{Battaglia_2017,Amodeo2021,Vavagiakis2021};
we consider the Dark Energy Spectroscopic Instrument's photometric Luminous Red Galaxy sample \citep[LRG;][]{Zhou2023LRG}, whose measured kSZ signal \citep{Hadzhiyska2024photoz, RiedGuachalla2025, Roper2025} is reproduced by the strongest feedback ({\eightsigma}) FLAMINGO simulation \citep{McCarthy2025,Siegel2025flamingo,Bigwood2025allthesims}.
Therefore, any deviation of the measured tSZ signal from the simulations likely arises from differences in the gas temperature. 

The galaxy sample and CMB maps are presented in Section~\ref{sec:data}.
The tSZ stacking measurements and modeling of astrophysical contaminants are described in Section~\ref{sec:tsz}.
We outline the simulation forward model in Section~\ref{sec:forward_model} and present our results in Section~\ref{sec:results}.
We discuss our findings in Section~\ref{sec:discussion} and conclude in Section~\ref{sec:conclusions}.

\section{Data}
\label{sec:data}

We measure the tSZ effect around the photometric DESI LRG sample (Section~\ref{sec:desi}) with the ACT~DR6 single-channel temperature maps (Section~\ref{sec:temp_maps}). The overlap of these data is shown in Appendix~\ref{appendix:ggl}, alongside the HSC lensing survey.

For comparison, we also consider the map-level tSZ estimates of \cite{Coulton2024}\footnote{\url{https://lambda.gsfc.nasa.gov/product/act/act_dr6.02/act_dr6.02_nilc_get.html}}; see Section~\ref{sec:depr_maps}.
{\CY} maps are powerful tools for studying the auto- and cross-correlations of the tSZ effect \citep[e.g.,][]{Ma2015,Pandey2019,Pandey2022,Efstathiou2025,Raghunathan2026}. 
Unlike the photometric approach, map-level methods are agnostic to the particular galaxy sample of interest \citep[e.g.,][]{Bennett1992, Maino2002, Cardoso2008, Remazeilles2011, FernandezCobos2012, Hurier2013}. 
This complicates the modeling of astrophysical contaminants because the redshifts and properties of the dust and radio sources are unknown. 

\subsection{DESI Luminous Red Galaxies}
\label{sec:desi}

The photometric sample of DESI LRGs was selected from the DESI Legacy Imaging Surveys Data Release 9 \citep{Dey2019}, using the $g$, $r$,
and $z$ optical bands and the WISE W1 band \citep{Zhou2023LRG}.
DESI spectroscopically calibrated photometric redshifts are publicly available from \cite{Zhou2023}, with median uncertainties below $0.02 \times (1+z)$.
We adopt the DR9 photometric redshift estimates because of the more uniform imaging; our results are consistent if we instead use the DR10 redshifts.
Photometric stellar mass estimates using a random forest algorithm are publicly available from \cite{Zhou2023}.

In this work, we divide the photometric LRG sample into four redshift bins: $z=(0.4, 0.54)$, $(0.54, 0.71)$, $(0.71,
0.86)$, $(0.86, 1.02)$, following \cite{Zhou2023}.
Each redshift bin is divided into narrow stellar mass bins.
To ensure sufficient signal-to-noise ratio in the tSZ measurement, we limit our analysis to $\log_{10}[M_\star / M_\odot]>11.25$.
The redshift and stellar mass distributions are presented in Appendix~\ref{appendix:ggl}.

\subsection{ACT CMB Temperature Maps}
\label{sec:temp_maps}

ACT \citep{Fowler2007} DR6 covers $19,000$ square degrees of the sky with $0.5'$ pixels to a median depth of $10~\mu \mathrm{K}$~arcmin with three channels \citep{Naess2025}: $77$--$112$ GHz, $124$--$172$ GHz,
and $182$--$277$ GHz, referred to as 90, 150, and 220~GHz, respectively. In our fiducial analysis, we consider the three co-added DR6 single-channel day-night temperature maps \citep{Naess2020, Naess2025}\footnote{\url{https://lambda.gsfc.nasa.gov/product/act/act_dr6.02/act_dr6.02_maps_coadd_get.html}}; our results are consistent if we instead use the night-only maps.

For each galaxy sample, we measure the stacked photometry from the single-channel maps and model the SED with three components: tSZ, dust, and radio emission \citep[e.g.,][]{Greco2015, Soergel2017, Vavagiakis2021, Meinke2021}; see Section~\ref{sec:sed_modeling}.
The stacked photometry method reduces systematic uncertainties in the modeling of astrophysical contaminants: the mean redshift of the galaxy sample is known, and uncorrelated foreground contaminants are mitigated with background subtraction.

The beam FWHMs of the three channels are $2.1'$, $1.4'$, and $1.0'$ for 90, 150, and 220~GHz, respectively.
For ease of modeling and to facilitate comparison with prior {\CY} studies, we convolve the maps to a common Gaussian beam of $1.6'$ FWHM. 
This convolution introduces noise for the 90~GHz map, which has a $2.1'$ beam, but this effect is suppressed by stacking.

Following \cite{Coulton2024} and \cite{Liu2025}, bright sources in the ACT maps are masked: point sources detected at $>5\sigma$ significance and extended sources at $>10\sigma$;
this masking only marginally affects the stacked photometry.
All tSZ clusters detected at $S/N>6$ are masked \citep{Aguena2026}.
Masking bright sources reduces the variance in our measurement and removes potential foreground contaminants;
we explore the significance of this masking in Section~\ref{sec:forward_model}.

\subsection{ACT+Planck Deprojected {\CY} Maps}
\label{sec:depr_maps}

The \cite{Coulton2024} {\CY} maps are derived from ACT and Planck data with the internal linear combination (ILC) method \citep{Bennett1992}.
The ILC method models the observed CMB data as a linear combination of underlying sources (e.g., the primary CMB anisotropies, the tSZ effect, and dust emission), with assumed models for the frequency dependence of each component. \cite{Coulton2024} do not model the radio contribution but do mask bright point sources;
by stacking the {\CY} maps on Galactic radio sources from the FIRST catalog, \cite{Battaglia2026} find the radio contamination in the maps is likely small.
The spatial dependence of the components is represented with wavelet kernels \citep[``needlets'';][]{Delabrouille2009}.
Before modeling, the ACT and Planck data are convolved to a common $1.6'$ Gaussian beam.

Because the shape of the dust SED is uncertain, \cite{Coulton2024} repeat the component separation process for a range of modified black-body SEDs (Equation~\ref{eqn:cib_sed}), spanning temperature $T_\mathrm{dust}=10.7$--$24$~K and spectral index $\beta=1.0$--$1.8$; see Section~\ref{sec:contaminants} for discussion.
The observed-frame dust SED is expected to vary across the sky---dust emission is sourced by galaxies at a wide range of redshifts ($z \lesssim 6$), and the dust properties evolve with redshift and galaxy type \citep[e.g.,][]{Chiang2025}---so assuming a single dust SED can leave significant residuals in the {\CY} map.
Following \cite{Chluba2017}, \cite{Coulton2024} include first-order Taylor expansions of the modified black-body SED around $\beta$ and/or $T_\mathrm{dust}$.
In this work, we focus on the CIB$+d\beta$ maps because they span the greatest range of {\CY} \citep[e.g.,][]{Liu2025}.

\section{tSZ Effect Measurements}
\label{sec:tsz}

To isolate the tSZ effect from astrophysical contaminants, we model the stacked SED with three components: tSZ, dust emission, and radio sources.
Because the photometry is background-subtracted (Section~\ref{sec:stacking}), the dust and radio emission arise from the targeted galaxies, as opposed to uncorrelated line-of-sight emission.
We can therefore leverage our knowledge of the galaxy sample, particularly the galaxies' redshifts and radio flux constraints, in contrast to {\CY} maps, for which the redshifts of the contaminants are unknown.

In Section~\ref{sec:stacking}, we describe the stacked aperture photometry measurements from the single-channel ACT maps.
The model components are introduced in Section~\ref{sec:contaminants}, and in Section~\ref{sec:sed_modeling} we describe the multi-frequency fitting procedure to isolate the {\CY} signal from astrophysical contaminants.

\begin{figure*}[t!]
\includegraphics[width=\textwidth]{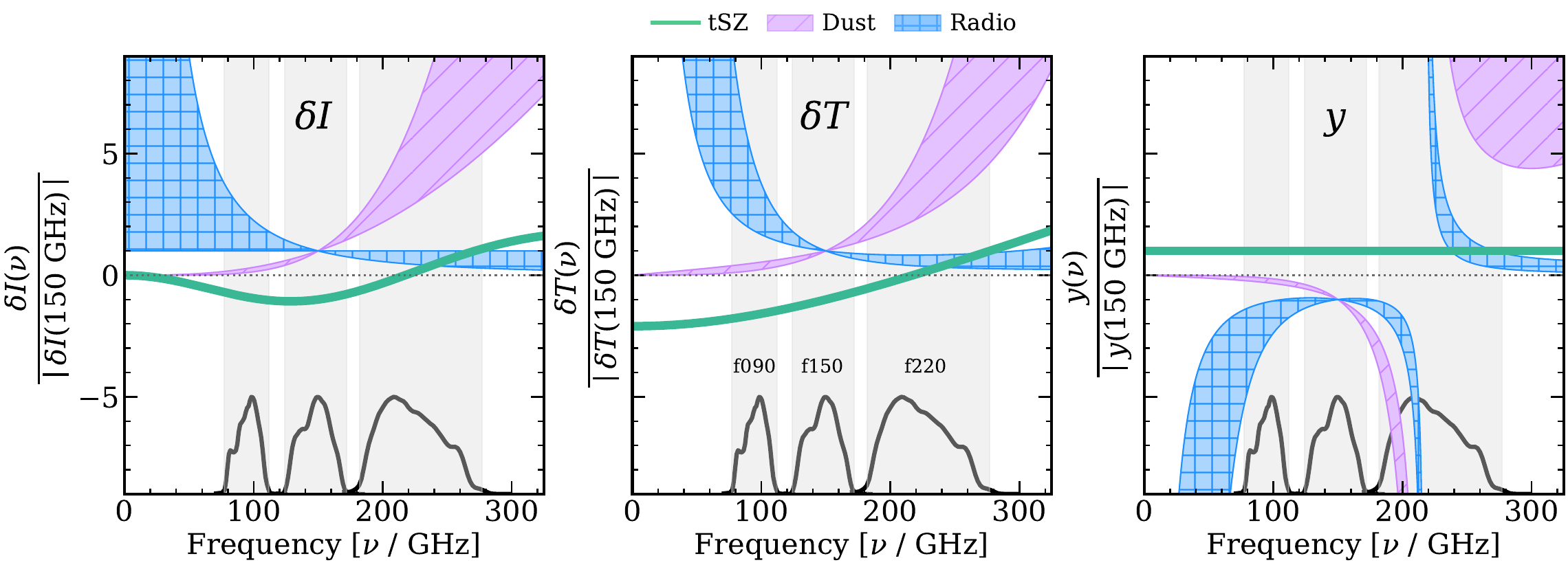}
\caption{
Normalized frequency dependence of the tSZ effect (green line), dust emission (purple, dash-hatched), and radio point sources (blue, cross-hatched).
The left (center) panel shows the induced flux (temperature) perturbation, relative to the primary CMB.
The rightmost panel shows the apparent {\CY} signal---i.e., converting the temperature fluctuations to $y$ assuming the fluctuations are sourced by the tSZ effect.
The dust SED is treated as a modified black-body, and
the radio SED is approximated as a power law.
To represent the uncertainty in the shape of the SEDs, we present the dust SED for $\beta \in [1,3]$ (see Equation~\ref{eqn:cib_sed}) and show the radio SED for $\alpha\in[-2,0]$ (see Equation~\ref{eqn:radio_sed}).
The frequency responses of the three ACT channels are shown in black (with arbitrary normalizations). 
}
\label{fig:frequency_dependence}
\end{figure*}

\subsection{Stacking}
\label{sec:stacking}

For a sample of galaxies located at $\{ {\bm \theta}_1, \dots, {\bm \theta}_i \}$ on the sky, we measure the stacked signal from a map $M(\bm{\theta})$ with compensated aperture
photometry (CAP) filters.
We consider the $90, 150,$ and $220$~GHz ACT single-channel temperature maps, as well as the \cite{Coulton2024} deprojected {\CY} maps.
The CAP filter sums the temperature fluctuations within $\theta_d$ of a galaxy's position ${\bm \theta}_i$ and then background-subtracts the fluctuations outside $\theta_d$;
this filter has the advantage of nulling all modes larger than $\theta_d$.
For the $i$th galaxy, the filtered signal is
\begin{align}
    \label{eqn:CAP}
    \mathcal{M}_i &= \int d^2\theta M(\theta)  W_i(\theta),\\
    W_i(\theta) &= \begin{cases}
        1 & \text{if } \theta < \theta_d\\
        -1 & \text{if } \theta_d \leq \theta \leq \sqrt{2}\theta_d\\
        0 & \text{if } \theta > \sqrt{2} \theta_d
    \end{cases},
\end{align}
where $\theta$ is the angular separation from the galaxy position ${\bm \theta}_i$.
We then take the mean of $\{\mathcal{M}_1,\dots,\mathcal{M}_i\}$ to yield the stacked signal.
To avoid contamination of the stacked profiles, we omit all galaxies within $15'$ of a masked pixel (see Section~\ref{sec:temp_maps}).

To probe the radial dependence of the tSZ signal, the CAP filter is calculated for $\theta_d=0.8'$--$6.4'$.
Because the filter is cumulative, the different radial bins are covariant, particularly at larger radii. 
The covariance matrix for a given galaxy sample is estimated via jackknife resampling.

\subsection{SED Model Components}
\label{sec:contaminants}

The microwave sky is illuminated by the primary CMB, the tSZ and kSZ effects (CMB secondaries), and astrophysical foregrounds: Galactic and extragalactic dust and radio point sources.
The background subtraction and stacking remove the primary CMB anisotropies, the kSZ effect, and uncorrelated foregrounds, but dust and radio emission from the targeted galaxies persists. 
To isolate the tSZ signal, we therefore model the stacked SED from the ACT $90$, $150$, and $220$~GHz photometry with three components: tSZ, dust emission, and radio point sources.
In this section we introduce the components of the model.
The fitting process is presented in Section~\ref{sec:sed_modeling}, including how we limit model degeneracies by fixing the shape of the dust SED; also see Appendix~\ref{appendix:contaminants}.

It is common (and convenient) to express flux perturbations on the sky in terms of an apparent temperature shift in the CMB black-body.
In the small perturbation regime, the observed flux intensity is
\begin{equation}
    \label{eqn:CMB_expansion}
    I_\nu({\bm \theta}) \approx B_\nu( T_\mathrm{CMB} ) + \left. \delta T_\nu({\bm \theta}) \frac{ \partial B_\nu }{ \partial T} \right|_{T_\mathrm{CMB}},
\end{equation}
where $B_\nu$ is the Planck function and $\delta T_\nu ({\bm \theta})$ is a small apparent temperature fluctuation at a position ${\bm \theta}$ on the sky. 
The perturbation in specific intensity relative to a perfect black-body is 
$\delta I_\nu ({\bm \theta}) = \left. \delta T_\nu({\bm \theta}) \frac{ \partial B_\nu }{ \partial T} \right|_{T_\mathrm{CMB}}. $
Following this convention, the SED model is defined in terms of $\delta T$.

The first component of the model is the tSZ effect, which imprints an apparent temperature shift in the CMB of 
\begin{equation}
\label{eqn:tsz}
\frac{\delta T_{\nu, \mathrm{tSZ}} ({\bm \theta}) }{ T_\mathrm{CMB} } = f_\mathrm{tSZ}(\nu) y({\bm \theta}),
\end{equation}
where $y$ is the dimensionless {\CY} parameter
\begin{equation}
y({\bm \theta}) = \frac{k_\mathrm{B} \sigma_\mathrm{T}}{m_\mathrm{e} c^2 } \int \frac{d\chi}{1+z} n_\mathrm{e} (\chi{\bm \theta}, z) T_\mathrm{e} (\chi {\bm \theta},z).
\end{equation}
$k_\mathrm{B}$ is the Boltzmann constant, $\sigma_\mathrm{T}$ is the Thomson cross section, $m_\mathrm{e}$ is the electron mass, $c$ is the speed of light, $\chi$ is comoving distance to redshift $z$, $n_\mathrm{e}$ is the electron number density, and $T_\mathrm{e}$ is the electron temperature. 
The frequency dependence of the apparent temperature shift in the non-relativistic limit ($k_\mathrm{B} T_\mathrm{e} \ll m_\mathrm{e} c^2$) is
\begin{equation}
\label{eqn:fnu_tsz}
f_\mathrm{tSZ}(\nu) = \frac{h \nu}{k_\mathrm{B} T_\mathrm{CMB}} \frac{ \exp \left[ \frac{h \nu}{k_\mathrm{B} T_\mathrm{CMB}} \right] +1 }{ \exp \left[ \frac{h \nu}{k_\mathrm{B} T_\mathrm{CMB}} \right] -1 } - 4.
\end{equation}
Relativistic electrons introduce a temperature dependence to $f_\mathrm{tSZ}$ \citep{Rephaeli1995, Itoh1998, Challinor1998,Chluba2012}; however,
for group-mass halos, the relativistic correction is small and subdominant to systematic uncertainties from foreground contamination.

The second model component is dust emission. 
UV photons from stars are absorbed and subsequently re-emitted by the surrounding dust.
At $10$--$10^4$~GHz the dust SED is well approximated as a modified black-body \citep{Blain2002, Blain2003, Casey2014}, for which the luminosity at frequency $\nu$ is given by
\begin{align}
    \label{eqn:cib_sed}
    L_{\mathrm{dust}}(\nu) &= L_\mathrm{bol, dust} \frac{\Phi_{\mathrm{dust}}(\nu)}{\int \Phi_{\mathrm{dust}}(\nu) d \nu},\\
    \Phi_{\mathrm{dust}}(\nu) &\propto \frac{\nu ^{3+\beta} }{ \exp\left[ \frac{h \nu}{ k_\mathrm{B} T_\mathrm{dust} } \right] -1 }, 
\end{align}
where $\Phi_{\mathrm{dust}}(\nu)$ is the rest-frame SED, $T_\mathrm{dust}$ is the dust temperature, and $\beta$ is the spectral index \citep{Draine1984,Meny2007}.
For a galaxy at redshift $z$, the observed-frame flux is
\begin{equation}
    \label{eqn:cibLtoF}
    S_{\mathrm{dust}}(\nu) = \frac{L_{\mathrm{dust}}(\nu \times[1+z])}{4 \pi \chi^2 (1+z)}.
\end{equation}
The specific intensity $\delta I_\mathrm{\nu, dust} = S_{\mathrm{dust}}(\nu) / \delta \Omega$ is related to an apparent CMB temperature fluctuation by Equation~\ref{eqn:CMB_expansion}, where $\delta \Omega$ is the solid angle.

The properties of the dust SED remain uncertain.
The dust temperature distribution is potentially broad \citep{Meisner2015}, and the dust properties are found to vary as a function of galaxy type and redshift \citep[e.g.,][]{Chiang2025}.
Observations at $10$--$10^4$~GHz indicate $T_\mathrm{dust} \approx 20$--$60$~K and $\beta \approx 1$--$2$ \citep[e.g.,][]{Dunne2000,Chapman2005,Amblard2010,Boselli2012,Shang2012,Addison2013,Planck2014CIB}.
In the ACT frequency range, $\beta$ and $T_\mathrm{dust}$ are degenerate \citep{McCarthyHill2024}.
We therefore fix $T_\mathrm{dust}=24$~K throughout this work, without loss of model flexibility.

At lower frequencies ($\lesssim 50$~GHz), Anomalous Microwave Emission (AME) is relevant, likely powered by spinning and/or magnetized dust grains \citep{Draine1998,Draine1999}.
However, for the ACT frequency range, AME is expected to be subdominant \citep{Thorne2017}.

The third component of the SED model is radio point source emission, primarily powered by synchrotron radiation from AGN.
The population of radio AGN is broadly divided into flat-spectrum ($L_\mathrm{Radio} (\nu) \propto \nu ^{\gtrsim -0.5}$) and steep-spectrum ($\propto \nu ^{\lesssim -0.5}$) sources, likely reflecting the orientation of the AGN with respect to the observer \citep[e.g.,][]{Urry1995}.
The average SED for the radio sources in the LRG sample is uncertain,
because the observed distribution of spectral indices $\alpha$ is broad \citep[][]{CalistroRivera2017,Zhong2025} and measurements of $\alpha$ are sensitive to frequency coverage and flux calibration.
Radio source counts from the South Pole Telescope (SPT) indicate average spectral indices of $\alpha\approx-0.7$ to $-0.6$ \citep{Everett2020}; however, the radio sources in the LRG sample may deviate from the bright millimeter sources detected by CMB telescopes.
From Planck data, \cite{Zhao2026} infer $\alpha \approx -0.7$ to $-0.2$ for the unWISE sample.
Free-free emission from the scattering of electrons off ions also contributes at low frequencies, with a spectral index of $\alpha \approx -0.1$ \citep{Bennett2013}. 
Given the limited spectral resolution afforded by the three ACT channels, we approximate the average radio SED as a power law
\begin{equation}
    \label{eqn:radio_sed}
    L_\mathrm{Radio} (\nu) = L_\mathrm{Radio, \nu_0} \left( \frac{\nu}{\nu_0} \right)^\alpha,
\end{equation}
where $L_\mathrm{Radio, \nu_0}$ is the rest-frame luminosity at $\nu_0$ and 
$\alpha$ is an effective spectral index, averaged over the population of radio emitters.
Exploring more complex radio SEDs with the Simons Observatory is warranted (discussed further in Section~\ref{sec:disc_robustness}).
The conversion from the rest-frame SED to a CMB temperature shift is analogous to Equation~\ref{eqn:cibLtoF}.

The frequency dependence of the three model components---tSZ, dust, and radio---is shown in Figure~\ref{fig:frequency_dependence}.
To reflect uncertainties in the shape of the dust and radio SEDs, we consider $\beta = 1$ to $3$ and $\alpha=-2$ to $0$;
we fix $T_\mathrm{dust}=24$~K because $\beta$ and $T_\mathrm{dust}$ are degenerate in this frequency range.
For frequencies below (above) $217$~GHz, the tSZ effect appears as a temperature decrement (excess) relative to $T_\mathrm{CMB}$.
Both the dust and radio sources increase the apparent temperature of the CMB.
The dust is dominant at higher frequencies, and radio sources are dominant at lower frequencies.

For each ACT channel, we convolve the SED model with the bandpass function.
The co-added temperature maps do not account for bandpass differences between the individual maps, introducing position- and scale-dependent variations in the bandpass center at the $0.5\%$ level \citep[][]{Naess2020};
this uncertainty is negligible for our analysis.
The ACT bandpass functions are shown in Figure~\ref{fig:frequency_dependence}.

\begin{figure*}[t!]

\gridline{ \fig{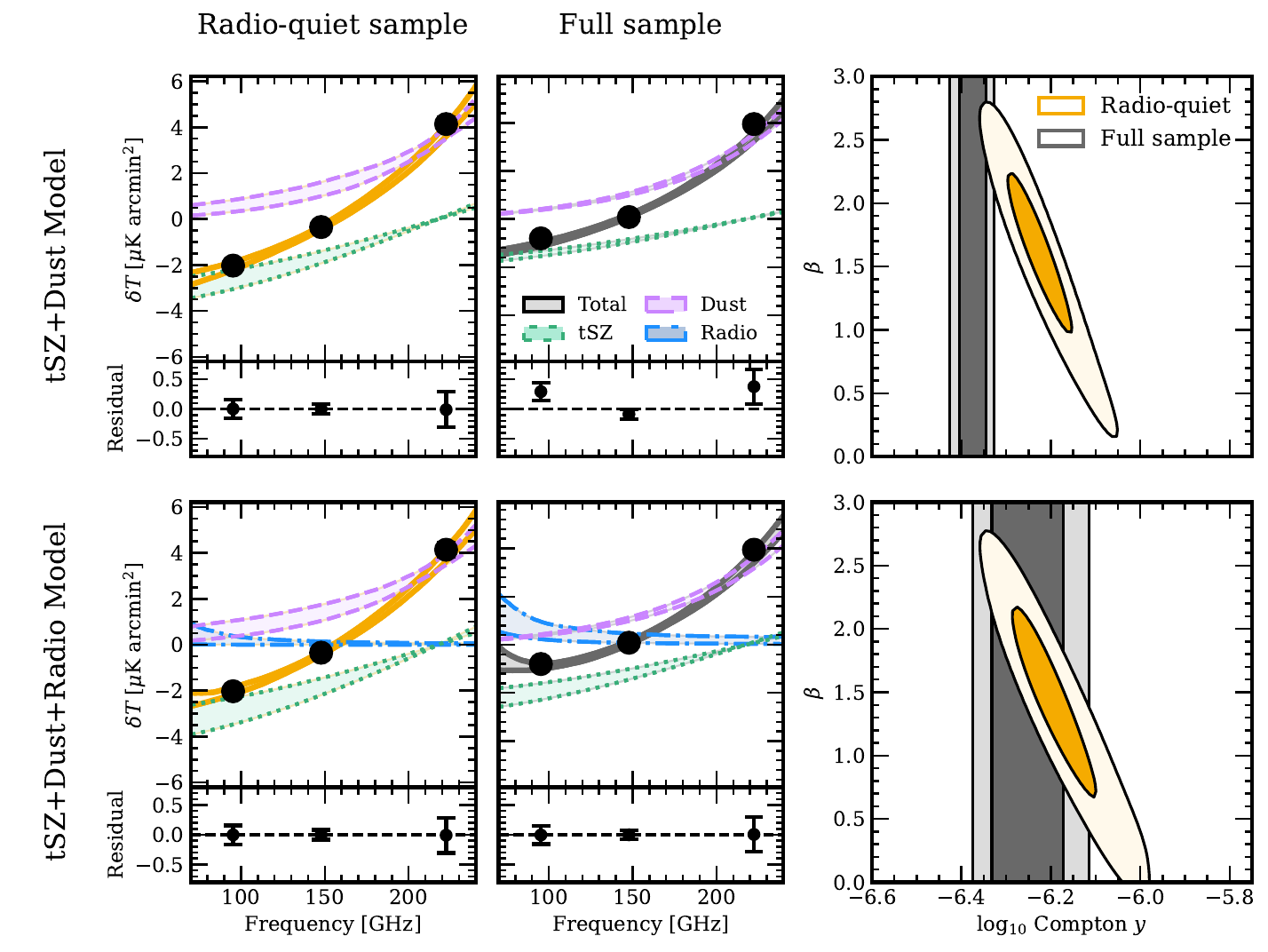}{\textwidth}{} }
\vspace{-2.em}
\caption{
The photometric LRG sample is well described by a joint tSZ, dust, and radio SED model.
\textit{Left} and \textit{middle} columns: stacked temperature fluctuations within the $1.6'$ aperture as a function of frequency for the radio-quiet LRG sample (left; yellow) and the full sample with no radio cut (middle; gray).
Both samples are at $z=0.54$--$0.71$ with $\log_{10} M_\star / M_\odot=11.25$--$12.0$.
The shaded bands show the $1\sigma$ posterior constraints on the SED model and its individual components.
The lower panels show the residuals.
For the radio-quiet sample, the dust spectral index $\beta$ is a free parameter; for the full sample, we fix $\beta=1.7$.
In the \textit{top} row, the SEDs are modeled with tSZ and dust components;
while the model successfully describes the radio-quiet SED, it fails to fit the full sample.
The \textit{bottom} row presents the joint tSZ, dust, and radio model, which successfully describes the radio-quiet sample, as well as the full sample.
\textit{Right column:} posteriors of the SED fits for each model.
}
\label{fig:radio_breakdown}
\end{figure*}

\begin{figure*}[t!]

\includegraphics[width=\textwidth]{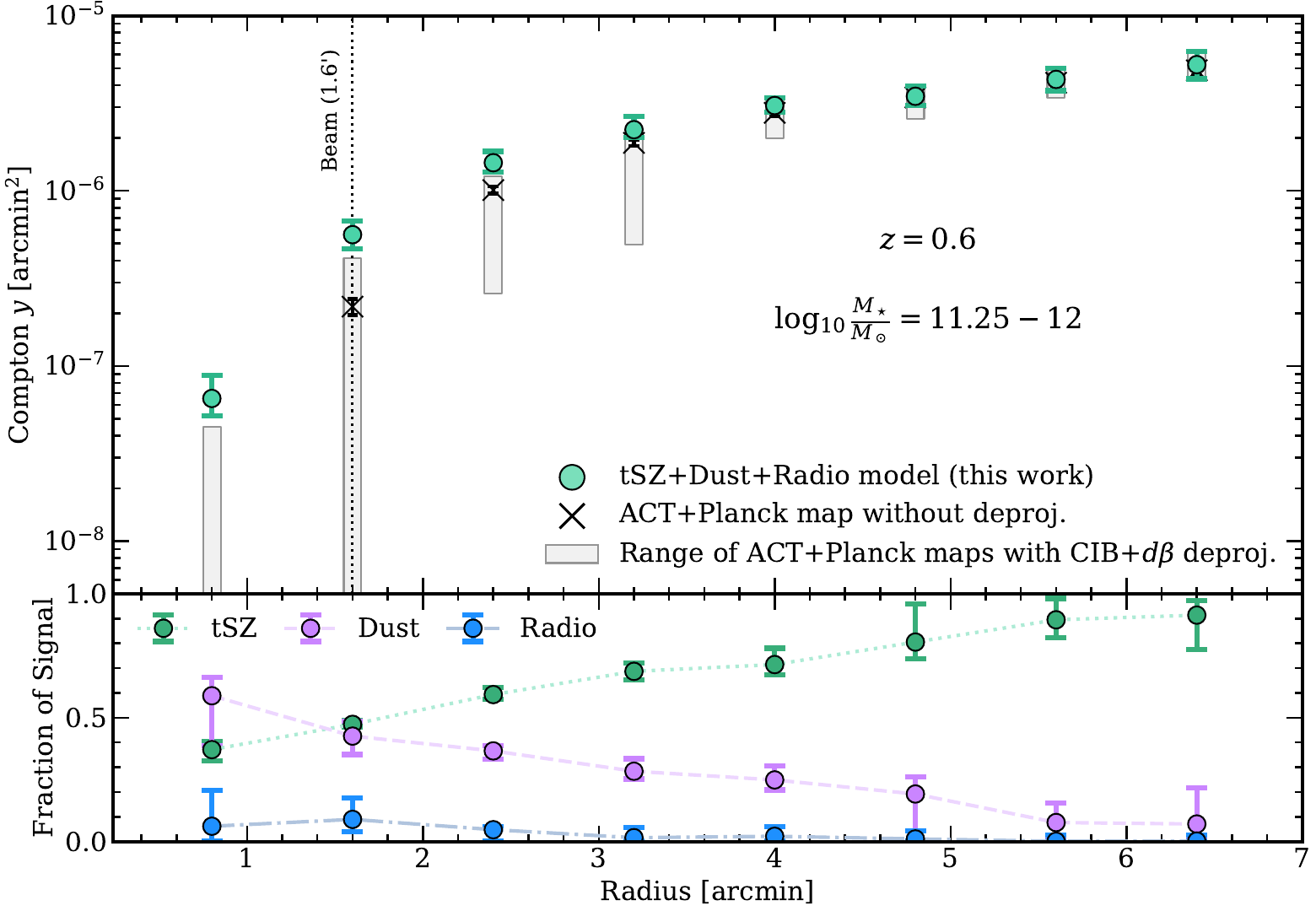}
\caption{
The astrophysical contaminants significantly bias tSZ constraints at small angular scales ($\lesssim 3'$). 
The SED model consistently recovers higher {\CY} values than existing deprojected ACT+Planck maps at radii where dust contamination is most important, while converging on large scales where the tSZ signal dominates.
For a representative galaxy sample, the top panel presents the {\CY} constraints and the bottom panel shows the relative contribution of the three model components;
we consider all LRGs at $z=0.54$--$0.71$ and $\log_{10} M_\star / M_\odot=11.25$--$12.0$.
The vertical dotted line marks the $1.6'$ aperture: the beam FWHM and the fiducial scale of our primary analysis.
The gray bars represent the range of stacked {\CY} from the CIB$+d\beta$ deprojected maps with $T_\mathrm{dust}=10.7$--$24$~K and $\beta=1.0$--$1.8$;
the stacked measurements from the ACT+Planck map without CIB deprojection are shown in black.
}
\label{fig:sed_demo}
\end{figure*}

\begin{figure}[t!]

\includegraphics[width=0.9\columnwidth]{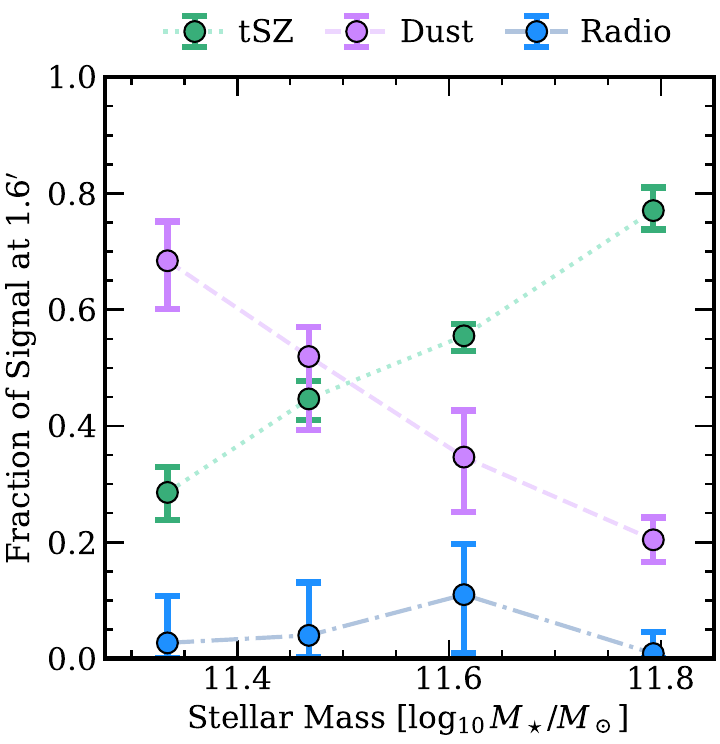}
\caption{
Fractional contribution of the SED model components---tSZ (green), dust (purple), and radio (blue)---to the stacked $150$~GHz signal within the $1.6'$
aperture (the FWHM of the beam) as a function of stellar mass.
We present the results for the $z=0.54$--$0.71$ redshift bin, which is representative of the LRG sample.
The dust emission dominates the lower mass bins, and the tSZ contribution rises steeply with stellar mass; the radio contribution remains low throughout.
}
\label{fig:contribution_vs_mass}
\end{figure}

\subsection{Model Fitting}
\label{sec:sed_modeling}

With only three channels, jointly constraining the tSZ, dust, and radio contributions is a highly degenerate problem.
To mitigate these degeneracies and restrict the model to physically plausible solutions, we fix the dust temperature $T_\mathrm{dust}$ and the spectral index $\beta$ for our fiducial analysis.
Even with these restrictions, degeneracies between the components persist. 
We therefore explore the posterior probability distribution with Markov Chain Monte Carlo (MCMC) sampling and conservatively focus our conclusions on the upper limit of {\CY}.
To further ensure conservative tSZ constraints, we model each galaxy sample and radial bin independently. 
We demonstrate that our conclusions are robust to a wide range of variations in Appendix~\ref{appendix:contaminants}, including adopting wide priors on the dust parameters.

To inform the dust properties ($T_\mathrm{dust}$ and $\beta$), we consider the radio-quiet subsample of LRGs.
For radio-quiet galaxies, the measured temperature fluctuations are dominated by the tSZ effect and dust emission,
allowing us to constrain the shape of the dust SED. 
We define radio-quiet galaxies as those with no source detected by the Very Large Array Sky Survey (VLASS) within $1.6'$ \citep{Lacy2020}.
VLASS spans declinations $>-40^\circ$ in the $2$--$4$~GHz range with $2.5''$ resolution.
We consider the epoch 2 quick look catalog \citep{Gordon2021}.
The $3 \sigma$ detection limit \citep[$360~\mu$Jy/beam;][]{Lacy2020} corresponds to $2.5~\mu$K~arcmin$^{2}$ for the $90$~GHz ACT channel, assuming a radio power-law index of $\alpha=-0.6$ \citep{Everett2020};
see Section~\ref{sec:disc_robustness} for further discussion.
The radio fraction depends weakly on redshift but increases with stellar mass, rising from $8\%$ in the lowest mass bin to $13\%$ in the highest.

As a representative example, Figure~\ref{fig:radio_breakdown} presents the stacked SED of radio-quiet galaxies at $z=0.54$--$0.71$ and $\log_{10} M_\star / M_\odot = 11.25$--$12.0$. 
The data are well fit by a joint tSZ and dust model; when a radio component is included, its amplitude is consistent with zero. 
We fix $T_\mathrm{dust}=24$~K \citep[e.g.,][]{Planck2014CIB,Lamperti2019,Chiang2025} and treat $\beta$ as a free parameter, inferring $\beta = 1.7 \pm0.5$.
Fixing $T_\mathrm{dust}$ alleviates the degeneracy with $\beta$ without diminishing the model's flexibility.
The inferred $\beta$ does not significantly change with redshift (Appendix~\ref{appendix:contaminants}).
The angular CIB power spectrum  \citep[e.g.,][]{Addison2013,Planck2014CIB,Yang2026} and CIB tomography \citep[e.g.,][]{Chiang2025} also favor $\beta \approx 1.7$; however, the precise value depends on the assumed dust temperature \citep[e.g.,][]{McCarthyHill2024}.
In this work, we assume $T_\mathrm{dust}=24$~K and $\beta=1.7$ for the fiducial SED model;
our conclusions are insensitive to the assumed dust properties (Appendix~\ref{appendix:contaminants}).

When radio galaxies are included in the LRG sample, the tSZ and dust ($T_\mathrm{dust}=24$~K and $\beta=1.7$) model no longer provides a good fit to the data (Figure~\ref{fig:radio_breakdown}).
In Appendix~\ref{appendix:contaminants}, we show that without a radio component, only unlikely dust properties ($\beta\gtrsim2.5$) can reproduce the data.
The inclusion of a radio component significantly improves the goodness-of-fit, although its contribution is found to be small ($\lesssim10\%$ at $150$~GHz). 
Recent studies have also found evidence of radio contamination in ACT and Planck data \citep{Zhao2026,Battaglia2026}.
The posteriors on {\CY} are consistent at the $1 \sigma$ level between the model with and without the radio component;
the inclusion of the radio component slightly raises the inferred {\CY}, reducing the discrepancy with simulations.
We adopt the joint tSZ, dust, and radio SED model for our fiducial analysis because it yields a more conservative simulation comparison and uses well-supported dust properties ($T_\mathrm{dust}=24$~K and $\beta=1.7$).

The {\CY} posteriors are consistent between LRG samples with and without radio galaxies (at the $1\sigma$ level); the radio-quiet sample is presented in Appendix~\ref{appendix:contaminants}.
Hereafter, we use the full LRG sample, regardless of radio properties.

The astrophysical contaminants are dominant for small apertures ($<3'$).
As a representative example, Figure~\ref{fig:sed_demo} presents the relative contribution of the tSZ, dust, and radio components as a function of radius, again for LRGs at $z=0.54$--$0.71$ and $\log_{10} M_\star / M_\odot = 11.25$--$12.0$.
The dust and radio contributions are both centrally concentrated; however, the dust still contributes at the $10\%$ level for the large $6.4'$ aperture.
In this work, we focus on smaller apertures, where the simulation predictions are most robust (Section~\ref{sec:disc_robustness}).

The relative contribution of the astrophysical contaminants also varies strongly with mass.
Figure~\ref{fig:contribution_vs_mass} presents the fractional contribution of the
tSZ, dust, and radio components to the $150$~GHz signal within the $1.6'$
aperture as a function of stellar mass.
In the lowest mass bin, the signal is dominated by the dust component, which sources $>60\%$ of the temperature fluctuation.
Because the tSZ effect scales strongly with halo mass, the tSZ component is dominant at higher masses;
however, the dust still contributes $\approx 20\%$ of the signal for the highest mass bin.
The radio contribution remains subdominant across the four mass bins.

We compare our tSZ constraints with the deprojected ACT+Planck {\CY} maps \citep{Coulton2024} in Figure~\ref{fig:sed_demo}.
Unlike our SED model, these maps are agnostic to the galaxy sample (i.e., the redshifts of the dust emitters are unknown); the maps also do not include radio deprojection (see Section~\ref{sec:depr_maps}).
For reference, we first consider the {\CY} map without deprojection (black crosses); 
as expected, the resulting {\CY} profile is significantly biased low, relative to our SED-derived result.
The CIB$+d\beta$ deprojected maps span a wide range of {\CY}, particularly for small apertures \citep[also see][]{Liu2025}.
The gray bars show the range of the {\CY} constraints across all the CIB$+d\beta$ deprojected maps ($T_\mathrm{dust}=10.7$--$24$~K, $\beta=1.0$--$1.8$).
A single deprojected map achieves similar precision to the SED model; however, the range of {\CY} spanned by the suite of deprojected maps is significantly greater than the statistical uncertainties. 
For the small apertures most affected by dust, all deprojected maps yield lower {\CY} than the SED model.
The CIB$+d\beta$ deprojected map that assumes the same dust properties as our SED fit ($T_\mathrm{dust}=24$~K and $\beta=1.7$) lies closer to our result than most of the maps, but it is still skewed lower (see Figure~\ref{fig:m_v_z_comp} and Section~\ref{sec:disc_robustness}).
On the largest scales, where the tSZ effect dominates, the deprojected maps and the SED model agree well.
Because the astrophysical contaminants are centrally concentrated, these discrepancies suggest the deprojected maps may harbor residual contamination on small scales.

For each galaxy sample, we constrain {\CY} by a joint tSZ, dust, and radio fit to the stacked photometry.
Our fiducial analysis assumes $T_\mathrm{dust}=24$~K and $\beta=1.7$.
We adopt wide uniform priors for the free parameters: the amplitude of the dust emission ($\log_{10}A_\mathrm{dust}$), the amplitude of the radio emission ($\log_{10}A_\mathrm{radio}$), the radio power-law index ($\alpha$), and the amplitude of the tSZ effect ($\log_{10}y$).

\section{Simulation Forward Model}
\label{sec:forward_model}

Comparing tSZ measurements with a simulation or analytic model is challenging.
The tSZ effect depends strongly on mass ($\propto M_\mathrm{halo}^{5/3}$ in the self-similar limit), so differences in halo mass between observation and theory can masquerade as differences in the tSZ effect. 
Because satellite galaxies reside in more massive halos than expected from their stellar mass and are miscentered from the host halo, the stacked tSZ effect is sensitive to the satellite fraction of the galaxy sample.
The tSZ effect measurement also includes contributions from halos along the line of sight (the two-halo effect).
To draw robust conclusions about the nature of baryon feedback and the state of the ICM, these effects must be taken into account when comparing observation and theory \citep[e.g.,][]{Planck2013tsz,Vikram2017,Popik2025,Kadir2026}.

To overcome these challenges, we perform a like-with-like comparison between stacked tSZ measurements and forward-modeled observations from the FLAMINGO simulations \citep{Schaye2023}.
For a given galaxy sample, we select a sample of simulated galaxies that best matches that sample's GGL measurement \citep{McCarthy2025};
GGL traces the total matter distribution from $0.1$ to $100$ comoving {\Mpch}, constraining the mean halo mass and satellite fraction of the galaxy sample.
We explore large variations in how the simulated galaxies are selected to ensure that our conclusions on the tSZ effect are robust.

The simulations and forward-modeled observations are introduced in Sections~\ref{sec:flamingo} and \ref{sec:forward_modeled_observables}.
The GGL measurements are briefly described in Section~\ref{sec:ggl}.
The like-with-like samples are outlined in Section~\ref{sec:selecting_simulated_samples}.

\subsection{FLAMINGO Simulations}
\label{sec:flamingo}

The FLAMINGO suite includes $16$ hydrodynamical simulations of varied resolution, box size, subgrid modeling, and cosmology \citep{Schaye2023}.
We consider the $1$~Gpc$^3$ intermediate-resolution simulations
($m_{\mathrm{gas}} = 1.09\times10^9$~{\Msol}), which include $2 \times 1800^3$ gas and dark matter particles and $1000^3$ neutrino particles.

Star formation, stellar evolution, radiative cooling, and AGN are treated via subgrid physics.
The strength of baryon feedback is primarily regulated by the AGN model.
The fiducial simulation employs radiative AGN feedback following \cite{Booth2009}.
The feedback energy is stored until it can heat the nearest gas particle by $\Delta T_\mathrm{AGN}$; releasing the energy each timestep would lead to numerical overcooling \citep{Dalla2012}.
Higher values of $\Delta T_\mathrm{AGN}$ correspond to less frequent but more powerful AGN outbursts.

The subgrid physics were calibrated to match the redshift-zero stellar-to-halo mass relation and the gas mass fractions of groups and clusters \citep{Kugel2023};
the gas fraction data were assembled from a collection of pre-eROSITA surveys, with limited corrections for selection effects.
The fiducial calibrated simulation successfully reproduces the galaxy stellar mass function, the central black hole--stellar mass relation, the cosmic star formation rate density, and cluster scaling relations \citep{Schaye2023,Braspenning2024}.
Simulation variants with stronger (weaker) baryon feedback were produced by calibrating to gas mass fractions artificially shifted down (up) by $N\sigma$, where $\sigma$ is the observational uncertainty on the mean gas fraction relation; we consider the fiducial and strongest feedback variants: $N=0$ and $-8$, respectively.
While all subgrid parameters were recalibrated simultaneously, the gas properties are most sensitive to the AGN subgrid model.

A like-with-like comparison between FLAMINGO and the SDSS/DESI+ACT kSZ effect profiles \citep{Schaan2021,Hadzhiyska2024photoz,RiedGuachalla2025} found that the strongest feedback simulation ({\eightsigma}) described the data best \citep{McCarthy2025,Siegel2025flamingo,Bigwood2025allthesims};
the strongest feedback simulation is also in better agreement with the gas mass fraction measurements of X-ray detected clusters from the first release of the eROSITA All-Sky Survey catalog \citep{Bulbul2024,Kovac25,Siegel2025flamingo}.
However, the strongest feedback simulation has also been shown to deviate from observed cluster scaling relations \citep[i.e., the $L$--$T$ relation;][]{Braspenning2024,Eckert2026};
the moderate feedback FLAMINGO variants (e.g., $f_\mathrm{gas}~-4\sigma$) are more consistent with the cluster scaling relations but overpredict the kSZ signal.

\begin{figure*}[t!]

\gridline{ \fig{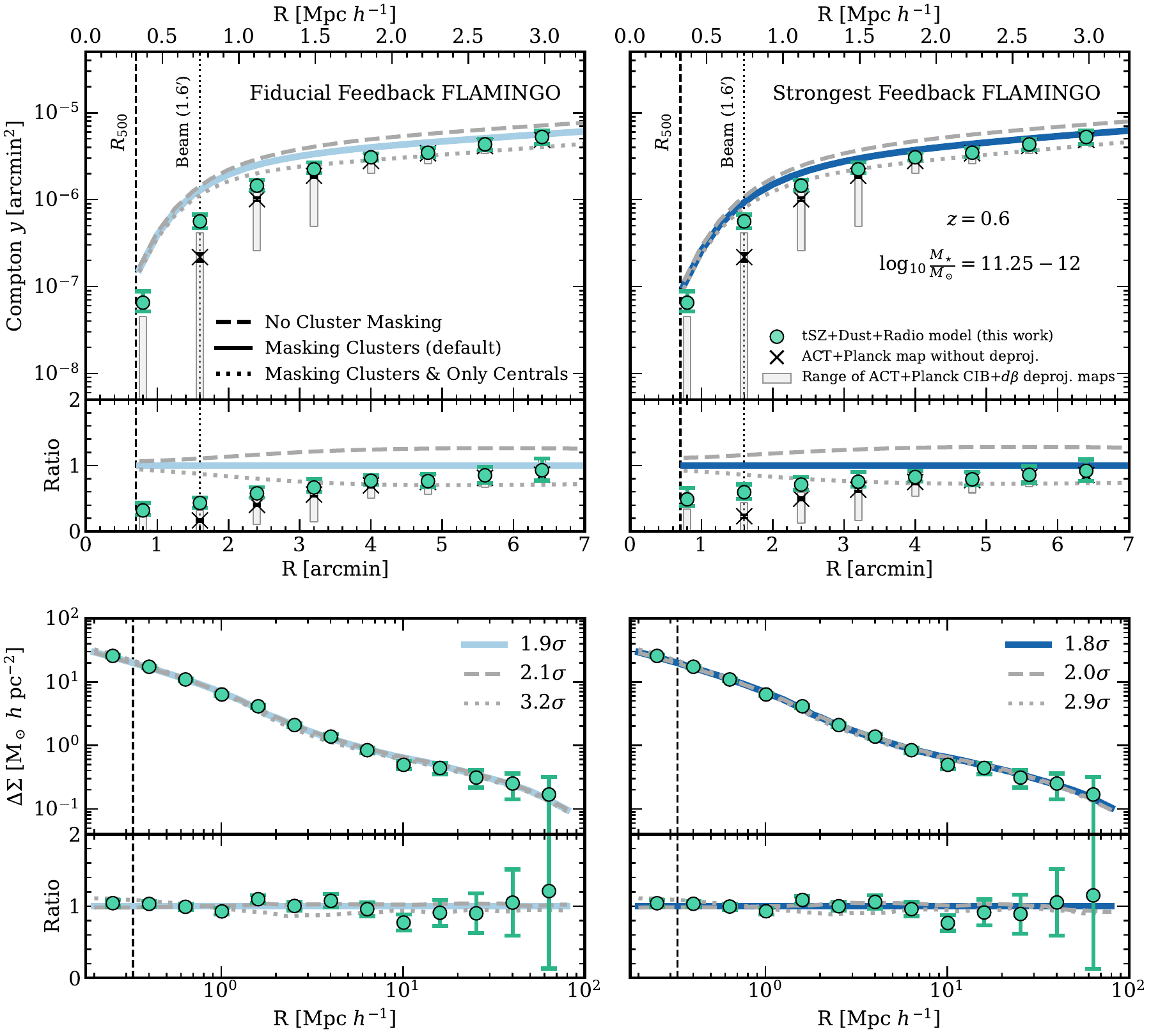}{\textwidth}{} }
\vspace{-0.6cm}
\caption{
Like-with-like comparison of the stacked {\CY} profile with the FLAMINGO simulations, for the same sample of LRGs as Figure~\ref{fig:sed_demo}.
The left and right columns present the predictions from the fiducial ({\fidsim}) and strongest feedback simulations ({\eightsigma});
the data are identical between the two columns.
\textit{Top:} CAP {\CY} profile from our SED modeling (green).
The gray bars represent the range of stacked {\CY} from the CIB$+d\beta$ deprojected maps with $T_\mathrm{dust}=10.7$--$24$~K and $\beta=1.0$--$1.8$;
the stacked measurements from the ACT+Planck map without CIB deprojection are shown in black.
The GGL-calibrated predictions from the FLAMINGO simulations are shown as solid lines;
predictions without satellites and without the masking of massive clusters are shown as dotted and dashed lines, respectively.
The bottom panels show the ratio relative to the nominal simulated galaxy sample. 
The vertical dashed line corresponds to $R_{500}$, and the vertical dotted line marks the $1.6'$ aperture: the beam FWHM and the fiducial scale of our primary analysis.
\textit{Bottom:} The measured GGL profile (green) for the sample of DESI galaxies alongside the best-fitting FLAMINGO GGL profiles. 
For each simulation prediction, we report the number of standard deviations by which the model deviates from the observed GGL. 
The radii are comoving distances.
}
\label{fig:sim_profile_comparison}
\end{figure*}

\subsection{Forward-Modeled Observables}
\label{sec:forward_modeled_observables}

For a given selection of simulated galaxies at redshift $z$ with coordinates $\{ {\bm \theta}_1, \dots, {\bm \theta}_i \}$ on the sky, the stacked tSZ effect and GGL profiles are calculated from the FLAMINGO lightcones \citep[Appendix~A of][]{Schaye2023}.

For the $i$th galaxy at redshift $z$, the $\Delta \Sigma$ profile is measured from the total mass map (gas, dark matter, stars,
black holes, and neutrinos) closest in redshift to $z$;
in the redshift range of the LRG sample, the lightcones are output in $\Delta z = 0.05$ intervals.
The individual $\{ \Delta \Sigma_1, \dots, \Delta \Sigma_i \}$ profiles are then stacked in comoving radial bins between $0.2$ and $75$~{\Mpch}.
We measure the CAP filtered tSZ effect for the $i$th galaxy from the cumulative {\CY} map, i.e., the sum of the lightcones.
As in the data, the {\CY} map is convolved with a $1.6'$ Gaussian beam. 
The $\{ y_1, \dots, y_i \}$ measurements are stacked in angular bins between $0.5'$ and $8'$.

This stacking process forward models the effects of miscentering and two-halo line-of-sight contributions, and marginalizes over the mass distributions of the simulated sample.
To reflect the redshift distribution of the observed galaxy sample $n(z)$, this process is repeated for $z=0.4$--$1.1$, and we take the $n(z)$ weighted average of the $\Delta \Sigma(R)$ and {\CY} profiles.

\subsection{Galaxy-galaxy lensing}
\label{sec:ggl}

For each galaxy sample in our tSZ analysis, we also measure the GGL profile.
As light from distant background galaxies (sources) travels toward us, it is tangentially sheared by all the intervening foreground structures (lenses); this effect is known as galaxy-galaxy lensing (GGL). 
The GGL measurements constrain the mean halo mass and satellite fraction of the galaxy sample, informing our like-with-like comparisons to the FLAMINGO simulations.
We closely follow the methodology described in \citet{Lange2024} and \citet{Heydenreich2025}, briefly outlined below.

For a foreground lens at redshift $z_{\rm l}$ and comoving distance $\chi_{\rm l}$, the induced tangential shear on a background source at angular separation $\theta$ from the lens is 
\begin{equation}
    \gamma_\mathrm{t}(R = \theta \chi_\mathrm{l}) = \frac{ \Delta \Sigma(R) }{ \Sigma_\mathrm{crit} },
\end{equation}
where $\Delta \Sigma(R) = \bar{\Sigma}(<R) - \Sigma(R)$ is the excess surface density: the difference between the average surface density within $R$ and the projected surface density at radius $R$.
$\Sigma_\mathrm{crit}$ is the comoving critical surface mass density
\begin{equation}
    \Sigma_\mathrm{crit}^{-1} = \frac{4 \pi G}{c^2} \frac{ D_\mathrm{l} D_\mathrm{ls} }{ D_\mathrm{s} } (1+z_\mathrm{l})^2,
\end{equation}
where $D_\mathrm{l}$ and $D_\mathrm{s}$ are the angular diameter distances to the lens and source, respectively, and $D_\mathrm{ls}$ is the relative distance.

We measure GGL through a weighted cross-correlation of the shapes of background sources $\epsilon$ with the positions of foreground lenses.
At a given angular separation $\theta$ between lens and source pairs, the tangential shear estimator is
\begin{equation}
    \langle \gamma_\mathrm{t} (\theta) \rangle = \frac{ \sum_{ \mathrm{ls} } \epsilon_\mathrm{t} w_\mathrm{ls} } { \sum_\mathrm{ls} w_\mathrm{ls} },
\end{equation}
where the summation is over all lens--source pairs at angular separation $\theta$, $\epsilon_\mathrm{t}$ is the tangential ellipticity component of each source, and $w_\mathrm{ls}$ is the combined weight of each lens--source pair. 
Considering an ensemble of source galaxies with a calibrated redshift probability distribution, $n(z_\mathrm{s})$, the average comoving critical surface density is
\begin{equation}
   \overline{\Sigma}_\mathrm{crit}^{-1}(z_\mathrm{l}) = \frac{4\pi G (1+z_{\rm l})^2}{c^2}\int d z_\mathrm{s} n(z_\mathrm{s}) \frac{D_{\rm l}  D_{\rm ls}}{ D_{\rm s}}.
\end{equation}
The average excess surface mass density is then
\begin{equation}
    \label{eqn:raw_esd}
   \Delta \Sigma (R) = \frac{ \sum_\mathrm{ls} \epsilon_\mathrm{t} w_\mathrm{ls}  } { \sum_\mathrm{ls} \overline{\Sigma}_\mathrm{crit}^{-1}(z_\mathrm{l}) w_\mathrm{ls} }.
\end{equation}

The GGL signals of the DESI lenses are measured with \texttt{dsigma} \citep{Lange2022} using the HSC Y3 shear catalog as sources.
The covariance matrix for each measurement is estimated with a leave-one-out jackknife process.
Fits to the GGL signals are restricted to $<20$~comoving~{\Mpch}, based on the size of the jackknife patches.
The GGL measurements are presented in Figure~\ref{fig:grid_ggl} and described in Appendix~\ref{appendix:ggl}.

\subsection{Like-with-like simulated samples}
\label{sec:selecting_simulated_samples}

For each galaxy sample, we select a sample of simulated galaxies that best fits the observed GGL profile \citep[][]{McCarthy2025}.
This calibration ensures that the observed and simulated samples have the same mean halo mass and satellite fraction, while bypassing reliance on uncertain stellar mass estimates;
for the DESI LRG sample, systematic uncertainties in stellar population synthesis modeling correspond to $\gtrsim 0.1$~dex uncertainty in mean halo mass \citep{Siegel2025flamingo}.

We nominally select galaxies according to a log-normal distribution of stellar mass:
\begin{align*}
    p\left(\log_{10}\left[\frac{M_\star}{M_\odot}\right]\right) = \mathcal{N}\left(\log_{10}\left[\frac{M_0}{M_\odot}\right], \sigma\right),
\end{align*}
where $M_0$ is optimized by $\chi^2$ minimization against the GGL measurement.
For our fiducial analysis, we fix $\sigma=0.2$~dex, motivated by the level of systematic uncertainty in the stellar mass estimates and the width of the stellar mass bins \citep{Siegel2025flamingo}.
This selection function qualitatively matches the observed stellar mass distributions, and has previously been shown to produce excellent matches to the SDSS and DESI GGL measurements \citep{Siegel2025flamingo}.
To mirror the masking of ACT-detected clusters in the data, we omit all simulated galaxies residing in halos with mass above some maximum: $M_{500}>M_{500, \mathrm{max}}$.
Based on the mass distribution of the ACT-detected tSZ clusters (Appendix~\ref{appendix:tsz_clusters}), we adopt a threshold of $M_{500, \mathrm{max}}=10^{14.3}$~{\Msol} for our fiducial analysis; this threshold is conservative, because it is a stricter masking criterion than that applied to the data, yielding lower simulated tSZ predictions than a more lax cut.
We determine the best-fitting selection of simulated galaxies for each observed galaxy sample independently.

The like-with-like comparison process is outlined in Figure~\ref{fig:sim_profile_comparison}.
For the same representative sample of DESI LRGs presented in Section~\ref{sec:tsz} ($z=0.54$--$0.71$ and $\log_{10} M_\star / M_\odot = 11.25$--$12.0$), we present the inferred {\CY} profile and the GGL measurement.
We compare the data with the best-fitting selection of simulated galaxies from the fiducial ({\fidsim}) and strongest feedback ({\eightsigma}) FLAMINGO simulations.
The best-fitting galaxy samples successfully reproduce the GGL measurement on a wide range of scales.
For both the fiducial and strongest feedback simulations, the predicted {\CY} is greater than the measurement on small scales ($<3'$).
The strongest feedback simulation provides a closer match to the tSZ data but still overpredicts {\CY}.
The agreement between the simulations and the data is better on larger scales; however, we caution that the simulation predictions are more sensitive to the satellite fraction and the masking of massive clusters at $\gtrsim3'$ (discussed below).

Because the tSZ signal scales strongly with halo mass ($\propto M_\mathrm{halo}^{5/3}$), the stacked tSZ signal is sensitive to the shape of the halo mass distribution, particularly the high-mass tail.
To explore the possible range of simulation predictions, we consider variants of the GGL calibration that each change the number of massive halos in the simulated sample: centrals only versus centrals and satellites, and with or without masking of massive clusters. 
Figure~\ref{fig:sim_profile_comparison} presents the centrals-only sample (dotted) and the sample without cluster masking (dashed) against the data.
The GGL calibration is repeated for each variant.
The centrals-only sample includes fewer massive halos and therefore shifts the tSZ signal down.
This selection is unphysical for the DESI LRGs, and it struggles to reproduce the observed GGL;
however, it is an informative test for how sensitive the simulation predictions are to the satellite fraction.
The sample without cluster masking produces a higher {\CY} profile than the fiducial selection;
this selection is also an unrealistic variant because ACT-detected clusters were masked in the data.
Appendix~\ref{appendix:simulation_variants} presents the simulation variants as a function of mass and redshift, and also considers additional variants, including unrealistically narrow stellar mass distributions. 

On the smallest scales ($<3'$), the simulation predictions are only weakly affected by the exclusion of satellites or the masking of clusters;
because the {\CY} profile is measured with the CAP filter, the smaller scales are dominated by the one-halo term. 
Our primary analysis therefore focuses on the smaller $1.6'$ apertures, where the simulation predictions are more robust.

\begin{figure*}[t!]

\gridline{ \fig{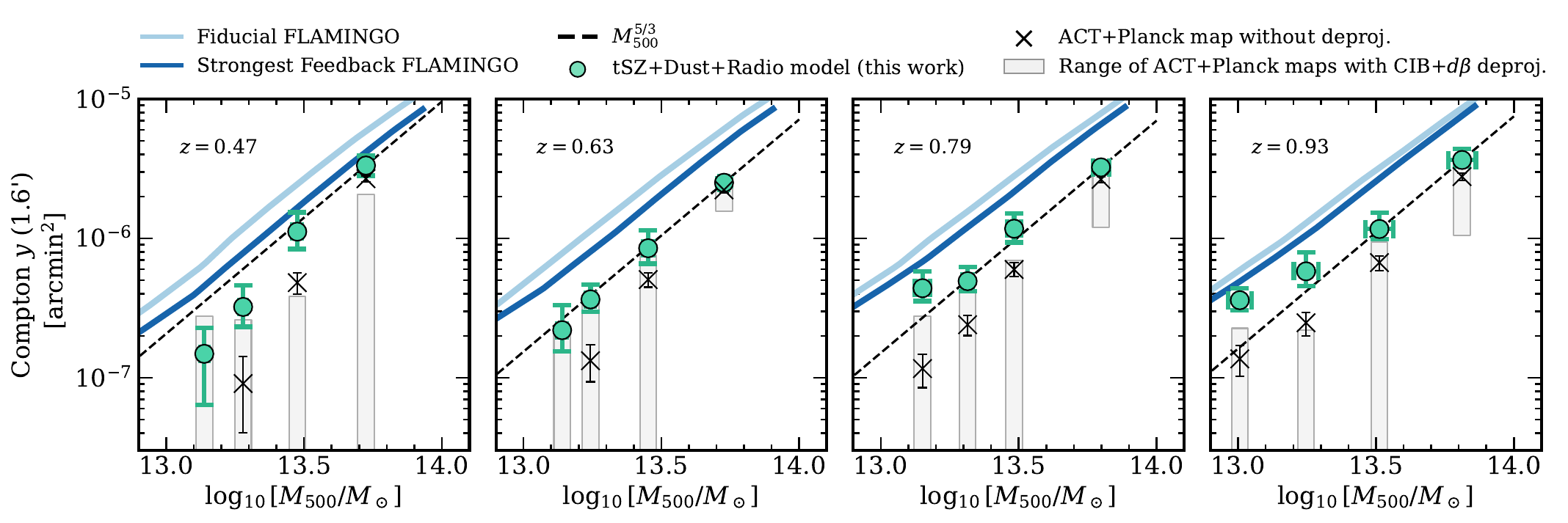}{\textwidth}{} }
\vspace{-0.6cm}
\caption{
Stacked {\CY} amplitude at $1.6'$ from the CAP filter as a function of mean halo mass.
Each column presents a different redshift bin; 
each redshift bin is divided into four stellar mass bins. 
The posteriors on {\CY} from the joint tSZ, dust, and radio model are shown in green.
The gray bars represent the range of stacked {\CY} from the CIB$+d\beta$ deprojected maps with $T_\mathrm{dust}=10.7$--$24$~K and $\beta=1.0$--$1.8$;
the stacked measurements from the ACT+Planck map without CIB deprojection are shown in black.
Predictions from the fiducial ({\fidsim}) and strongest feedback ({\eightsigma}) FLAMINGO simulations are shown as light and dark blue lines, respectively.
The dashed line presents the $M_{500}^{5/3}$ self-similar scaling with an arbitrary normalization. 
}
\label{fig:m_v_z_fiducial}
\end{figure*}

\begin{figure*}[t!]
\includegraphics[width=\textwidth]{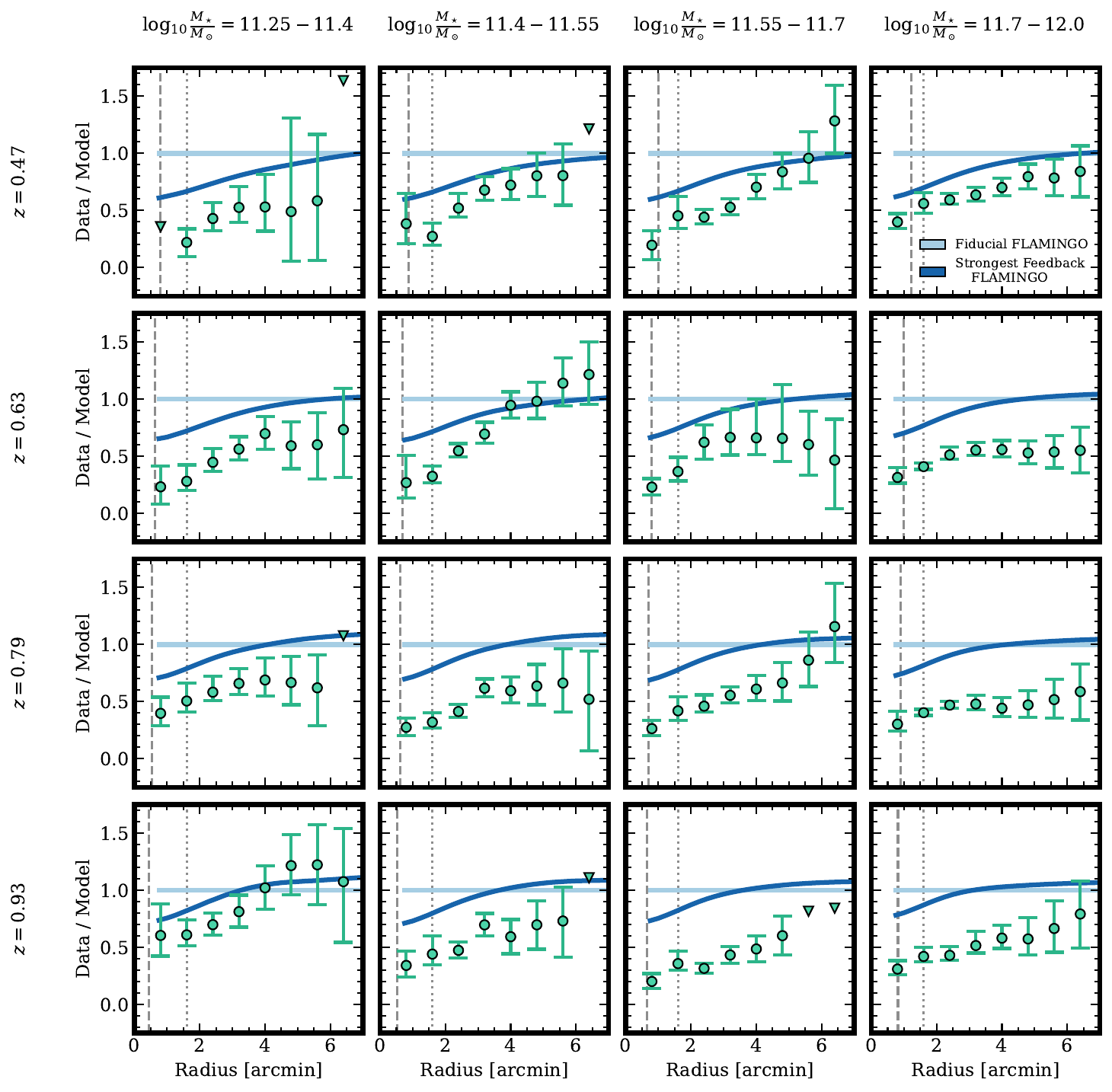}
\caption{
The {\CY} profiles (green) relative to the GGL-calibrated predictions of the fiducial ({\fidsim}; light blue) FLAMINGO simulation; the strongest feedback simulation ({\eightsigma}) is shown in dark blue.
Each row corresponds to a different redshift bin, and each column presents a different stellar mass bin.
The dashed vertical lines present $R_{500}$ for each sample, and the vertical dotted lines mark $1.6'$: the beam FWHM and the fiducial scale shown in Figure~\ref{fig:m_v_z_fiducial}.
Where the tSZ constraint is consistent with zero, we show the $2\sigma$ upper limit as a triangle.
}
\label{fig:grid}
\end{figure*}

\section{Results}
\label{sec:results}

We present new stacked measurements of the tSZ and GGL effects for bins in mass and redshift, and perform a like-with-like comparison with state-of-the-art hydrodynamical simulations. 
The galaxies are drawn from the photometric DESI LRG sample: $M_{500}=10^{13}$--$10^{14}$~{\Msol} and $z=0.4$--$1.0$.
The {\CY} profiles are measured from ACT~DR6 single-channel temperature maps with a joint tSZ, dust, and radio SED model.
The GGL signals are measured from the HSC~Y3 shear catalog.
The data are described in Section~\ref{sec:data}.

The tSZ profiles span $0.8'$ to $6.4'$ (i.e., $0.4$ to $3$~comoving {\Mpch} at $z=0.7$).
Our primary analysis focuses on the $1.6'$ aperture, which represents the smallest scale accessible without dropping below the beam size. 
At this scale, the simulation predictions are more stable to the effects of satellite galaxies and massive clusters, both of which can introduce modeling uncertainty at larger scales (see Figure~\ref{fig:sim_profile_comparison} and Appendix~\ref{appendix:simulation_variants}).
Although astrophysical contaminants are more significant at these small apertures, particularly at lower masses, we demonstrate in Section~\ref{sec:disc_robustness} and Appendix~\ref{appendix:contaminants} that our conclusions are robust to large variations in the SED fitting process, including omitting the radio component entirely and freeing the dust properties. 
Our fiducial SED model yields the highest {\CY} and is therefore a conservative choice for our simulation comparisons.

Figure~\ref{fig:m_v_z_fiducial} presents {\CY} within the $1.6'$ CAP aperture as a function of mean halo mass in four redshift bins.
The mean halo masses are derived from the GGL-based simulation comparisons.
The dashed line presents the $M_{500}^{5/3}$ self-similar scaling with an arbitrary normalization. 
Although the tSZ signal approximately follows the self-similar scaling, we caution that a power law is an oversimplification. 
Because {\CY} is measured at a fixed angular scale, the enclosed fraction of $R_{500}$ varies as a function of mass and redshift;
the two-halo contribution also depends on mass and redshift \citep[e.g.,][]{Vikram2017}.

For comparison, we also show the stacking results from the ACT+Planck {\CY} maps \citep{Coulton2024}.
The ranges of {\CY} spanned by the CIB$+d\beta$ deprojected maps are shown as gray bars; the map variants include $T_\mathrm{dust}=10.7$--$24$~K and $\beta=1.0$--$1.8$.
The deprojected maps consistently produce lower {\CY} than our SED-derived constraints. 
The discrepancy between the SED model result and the ACT+Planck maps is largest for small apertures and low masses, where astrophysical contaminants contribute the most (Section~\ref{sec:sed_modeling}).

For each galaxy sample, we perform a GGL-calibrated simulation comparison. 
The predictions are drawn from the sample of simulated galaxies that best fits the observed GGL profile.
As outlined in Section~\ref{sec:forward_model}, our primary selection assumes a log-normal stellar mass distribution ($\sigma=0.2$~dex), including both centrals and satellites, with the most massive clusters masked ($M_{500,\mathrm{max}}=10^{14.3}$~{\Msol});
the center of the stellar mass distribution is optimized to best fit the observed GGL profile.
We discuss variants of the simulation predictions in Section~\ref{sec:disc_robustness} and Appendix~\ref{appendix:simulation_variants};
as noted above, we focus on $1.6'$ apertures because they are weakly sensitive to the effects of satellites and massive clusters.
The simulation predictions from the fiducial ({\fidsim}) and strongest feedback ({\eightsigma}) FLAMINGO simulations are presented alongside the {\CY} measurements in Figure~\ref{fig:m_v_z_fiducial}.

The fiducial FLAMINGO simulation ({\fidsim}) overpredicts the {\CY} signal by $\gtrsim 2\times$ at $1.6'$ over the full mass and redshift range.
The strongest feedback ({\eightsigma}) simulation provides a better match to the data, approximately reproducing the higher-mass observations at $z=0.5$ but overpredicting the {\CY} signal elsewhere.
For most of the mass and redshift bins, the discrepancy between the simulations and the data is larger than the difference between the fiducial and strongest feedback simulations.
In Section~\ref{sec:disc_robustness}, we demonstrate that the deficit is robust to systematic uncertainties in the simulation predictions and {\CY} measurements; 
because these uncertainties dominate over the statistical errors, particularly for the simulation prediction, we refrain from quoting a formal significance for the deficit.
The discrepancies between the data and the simulations are even greater if the ACT+Planck {\CY} maps are used; 
the strongest feedback simulation deviates from all of the ACT+Planck {\CY} maps throughout the full mass and redshift range.

In Figure~\ref{fig:grid}, we zoom out and consider the full radial profile of the tSZ effect ($0.8'$--$6.4'$);
because the profile is cumulative, the radial bins are covariant.
The simulations consistently overpredict {\CY} within $3'$.
For some bins, particularly at $z=0.5$, the simulations reproduce the larger-aperture measurements; in others, they still overpredict {\CY}.
However, we caution that the simulation predictions are more sensitive to the effects of satellites and massive clusters on larger scales.
Given these systematic uncertainties, we defer detailed analysis of the radial dependence to future work.

\subsection{Robustness of the thermal pressure deficit}
\label{sec:disc_robustness}

Both the measurement and the theory prediction of the tSZ effect are subject to systematic uncertainties: the tSZ signal is contaminated by astrophysical foregrounds, and simulation predictions are acutely sensitive to the halo mass and satellite fraction of the galaxy sample. 
In this section, we review our steps to mitigate these uncertainties and bound the impact of residual systematics, demonstrating that the thermal pressure deficit result is robust.
We focus on establishing the existence of the deficit and leave the study of its potential mass and redshift dependence to future work.

\paragraph{\textbf{Astrophysical contamination.}}
Dust and radio emission ``fill in'' the tSZ temperature decrement, biasing {\CY} constraints low unless properly modeled. 
Contamination is most severe within small apertures and at low masses (Figures~\ref{fig:sed_demo} and \ref{fig:contribution_vs_mass}).

Our {\CY} constraints are stable against large variations in the SED fitting process.
The fiducial {\CY} constraints are derived assuming $T_\mathrm{dust}=24$~K and $\beta=1.7$.
In Appendix~\ref{appendix:contaminants}, we repeat the SED modeling with a more flexible dust component---$\beta\in[0,3]$---finding the {\CY} constraints shift \textit{down} by $<1\sigma$, increasing the apparent pressure deficit.
Because the shape of the dust SED is highly degenerate with the radio component, these fits are disfavored, but they reveal the potential range of {\CY} supported by the data.
Our conclusions are also insensitive to the radio component. 
If the radio component is omitted, the {\CY} constraints again shift down by $<1\sigma$.
The fiducial SED model therefore produces the highest {\CY} constraints, which still lie below the simulation predictions.

As an external check on our SED model, we test whether the inferred dust and radio properties are plausible for LRGs.
Because the luminosity of dust emission is proportional to the star formation rate (SFR)---$L_\mathrm{bol} \approx 10^{10} L_\odot \times \frac{\mathrm{SFR}}{M_\odot~\mathrm{yr}^{-1}}$ \citep{Kennicutt1998}---the inferred dust component should be consistent with the observed SFR of LRGs.
The SED fits yield $L_\mathrm{bol}\approx5\times10^{10}~L_\odot$, corresponding to $\mathrm{SFR} \approx 5~{M_\odot~\mathrm{yr}^{-1}}$, with statistical uncertainties on the order of $1~{M_\odot~\mathrm{yr}^{-1}}$. 
Publicly available \texttt{FastSpecFit}\footnote{\url{https://github.com/desihub/fastspecfit}} \citep{fastspecfit} modeling of the one-dimensional DESI spectra finds a mean SFR of $\approx 4~{M_\odot~\mathrm{yr}^{-1}}$ for the mass range considered in this work.
This agreement is encouraging; however, our SFR constraints are systematics-dominated, because the dust SED must be extrapolated beyond the ACT frequency range.
For the radio component, we consider external constraints from VLASS. 
By cross-matching the LRGs with the epoch 2 quick look catalog \citep{Gordon2021}, we derive a $3\sigma$ upper limit of $800~\mu$Jy on the galaxies' mean radio flux at $3$~GHz.
In Appendix~\ref{appendix:contaminants}, we demonstrate that our tSZ constraints are unchanged if we impose this $3 \sigma$ upper limit on the SED model.
Radio sources with flat SEDs could be undetected by VLASS but still be significant at ACT frequencies---the $3 \sigma$ VLASS limit corresponds to $20~\mu$K~arcmin$^{2}$ for $\alpha=0$;
it will therefore be informative to investigate the radio contribution with the Simons Observatory in the near future.
Together, these tests show that the dust and radio properties inferred by the SED model are consistent with external constraints.

For large apertures and high masses, where the tSZ signal dominates, our {\CY} constraints are consistent with the ACT+Planck {\CY} maps of \cite{Coulton2024}.
However, the map-derived {\CY} is systematically lower than our results where contamination is significant (Figures~\ref{fig:sed_demo} and \ref{fig:m_v_z_fiducial}).
These discrepancies suggest that the deprojected maps may harbor residual astrophysical contamination. 
The CIB$+d\beta$ deprojected map that assumes the same dust properties as our fiducial model agrees with our measurements better than most of the maps, but it still yields lower {\CY}.
Even for the same assumed dust properties, the models could differ in terms of the spatial distribution of dust or the effective redshift of the dust SED; the deprojected maps determine the effective redshift of the dust emission via the first-order expansion about $\beta$, while the SED model adopts the redshifts of the targeted galaxies.
The data also differ between the two methods: our stacked photometry isolates the dust emission correlated with the targeted galaxies, while the deprojected map must model that emission plus the uncorrelated CIB. 
Adopting the {\CY} maps would increase the apparent pressure deficit.

\paragraph{\textbf{Simulation predictions.}}
Comparing tSZ measurements with theory is complicated by the signal's strong scaling with halo mass.
We address this challenge by calibrating the selection of simulated galaxies against each sample's GGL profile.

The thermal pressure deficit is robust to even dramatic changes in our simulation comparison.
The fiducial selection draws centrals and satellites from a log-normal stellar mass distribution and, mirroring the masking of massive clusters in the tSZ measurement, omits simulated halos with $M_{500}>10^{14.3}$~{\Msol};
relaxing the cluster mask increases the discrepancy with the data.
This fiducial selection is designed to accurately reflect the observed galaxy samples: the width of the stellar mass distribution ($\sigma=0.2$~dex) is based on the level of systematic uncertainty in the stellar mass estimates \citep{Siegel2025flamingo}, and the selection has been shown to provide excellent agreement with the GGL measurements \citep{McCarthy2025}; also see Appendix~\ref{appendix:ggl}.
As an external check on the simulated galaxy samples, we compare the widths of the halo mass distributions with those predicted by the \textsc{UniverseMachine} mock catalogs \citep{Behroozi2019}---a model linking galaxy and halo assembly, using observational data---finding strong agreement (within $10\%$).

To illustrate the possible range of the tSZ predictions, we consider two extreme cases: narrowing the distribution to $\sigma=0.05$~dex and excluding satellites entirely; see Appendix~\ref{appendix:simulation_variants} for discussion.
Both variants suppress the stacked {\CY} signal by admitting fewer massive halos;
the resulting halo mass distributions are nearly half the width of the distributions predicted by the \textsc{UniverseMachine} mock catalogs \citep{Behroozi2019}.
Even these unphysical simulated samples do not fully resolve the deficit (Figure~\ref{fig:grid_comp}).
Because these variants are implausible descriptions of the DESI LRGs, the {\CY} signals from these tests are informative lower bounds.

\section{Discussion}
\label{sec:discussion}

Having established the robustness of the thermal pressure deficit, we now consider its physical origin and implications for other observables.

\subsection{Physical origins of the deficit}
\label{sec:disc_physics}

Because the tSZ effect constrains the thermal pressure---{\CY}~$\propto\int n_\mathrm{e} T_\mathrm{e}\,dl$---the discrepancy with the simulations could arise from differences in gas density and/or temperature.
It has been shown that the fiducial FLAMINGO simulation ({\fidsim}), which was calibrated to reproduce pre-eROSITA gas mass fractions \citep{Kugel2023}, overpredicts the kSZ signal of the DESI LRGs, indicating the simulated groups are too gas-rich \citep{Siegel2025flamingo,Bigwood2025allthesims}.
Therefore, the tSZ signal is expected to deviate from the fiducial simulation, even if the gas temperature is accurate.
However, the strongest feedback simulation ({\eightsigma}) reproduces the kSZ data, yet still overpredicts the tSZ signal.
Further gas depletion would strain the agreement with the kSZ data;
because the remaining tSZ discrepancy found in this work is typically larger than the difference between the fiducial and strongest feedback predictions, the required amount of additional depletion would be large.
The {\CY} deficit thus points to an overprediction of the gas temperature by $\approx 2\times$.

In hydrostatic equilibrium, the pressure gradient balances gravity: $\frac{d P_\mathrm{tot}}{dr}= - \frac{G M(r)\rho_\mathrm{gas}(r)}{r^2}$, where $M(r)$ is the enclosed mass and the total pressure $P_\mathrm{tot}$ consists of a thermal $P_\mathrm{th} \propto T_\mathrm{gas} n_\mathrm{gas}$ and a non-thermal $P_\mathrm{nt}$ component. 
Lowering the gas temperature by a factor of $\approx2$ would require significant deviations from hydrostatic equilibrium and/or non-thermal pressure contributions. 

Cosmological hydrodynamical simulations, which self-consistently follow the hierarchical growth of halos, including disturbances from mergers and accretion, find that most halos are close to hydrostatic equilibrium.  
In FLAMINGO, the median volume-weighted hydrostatic mass bias is $\lesssim 10\%$ at $M_{500}=10^{13}$~{\Msol} and near zero at $M_{500}\gtrsim 10^{14}$~{\Msol} \citep{Braspenning2025hydro};
the X-ray-weighted mass bias, which is relevant for observational studies of X-ray-inferred masses, is higher (at the $10\%$ level for groups and clusters).
Non-gravitational processes that act on relatively short timescales could drive the gas away from hydrostatic equilibrium.  For example, if the cooling time of the gas is comparable to, or shorter than, the free-fall time, no stable hydrostatic atmosphere can form \citep[e.g.,][]{Birnboim2003,Keres2005}.  
At the mass and radial scales under consideration here, such rapid cooling is not expected to be an important factor.  
Indeed, X-ray observations reveal that on $R_{500}$ scales of groups and clusters, the cooling time exceeds the Hubble time  \citep[e.g.,][]{Panagoulia2014,Hogan2017}; metal-dependent cooling of the gas is followed in modern cosmological simulations.
Energetic feedback events could also result in significant outflows that drive the gas away from equilibrium.  
However, the strong feedback variants in FLAMINGO, which are capable of reproducing the low gas fractions implied by recent kSZ and X-ray observations, are also near hydrostatic equilibrium, and still significantly overpredict the tSZ measurements presented here.
This lack of deviation from hydrostatic equilibrium due to feedback likely reflects that in FLAMINGO the low gas fractions of groups and clusters are not achieved by gas ejection from these halos, but by ejection from their $\sim$Milky-Way-mass, high-redshift progenitors \citep[e.g.,][]{McCarthy2011,Costello2026}.  
It remains an open question whether strong outflows could be relevant for the equilibrium of groups and clusters. 

In hydrostatic equilibrium, lowering the thermal pressure requires an increased non-thermal contribution. Below, we assess which sources of non-thermal pressure support could resolve the deficit: gas motions, non-equilibrium between electrons and ions, cosmic rays, and magnetic fields.

Hydrodynamical simulations self-consistently include non-thermal pressure support from gas motions.
For the FLAMINGO simulations, kinetic motions contribute moderately to the total pressure within $R_{500}$, falling from $15\%$ at $M_{500}=10^{13}$~{\Msol} to $5\%$ at $M_{500}=10^{14}$~{\Msol} \citep{Braspenning2025hydro}.
Kinetic contributions of $\approx 0.1 P_\mathrm{tot}$ are typical for hydrodynamical simulations at group masses, with the contribution increasing as a function of radius \citep[e.g.,][]{Lau2009,Battaglia2012,Nelson2014}.
The gas motions are primarily due to hierarchical structure formation, and feedback implementations only vary the kinetic pressure at the $10\%$ level \citep{Braspenning2025hydro}.
Observational constraints are typically limited to nearby clusters.
For the cool-core clusters Abell~2029 \citep{Xrism2025} and Perseus \citep{Zhang2026Perseus}, recent XRISM gas kinematics measurements find non-thermal pressure contributions of $\approx2.5\%$ and $\approx 10\%$, respectively;
\cite{Eckert2025coma} infer a $\approx10\%$ contribution for the Coma cluster but note the assumption of hydrostatic equilibrium may not hold in the post-merger phase.
Resolving the observed thermal pressure deficit with kinetic pressure alone, under the assumption of hydrostatic equilibrium, would require significantly more support from gas motions than indicated by current simulations and observations; other sources of non-thermal pressure support are likely required.

Non-equilibrium between electrons and ions is also unlikely to resolve the observed thermal pressure deficit.
Because ions are heavier, they carry more of the kinetic energy of the infalling gas than the electrons. 
If the timescale for electrons and ions to equilibrate via Coulomb collisions is long enough, the electron temperature can be significantly lower than that of the ions \citep{Fox1997,Ettori1998}.
The timescale of Coulomb collisions scales as $t_\mathrm{c} \propto T_\mathrm{e}^{3/2} / n_\mathrm{i}$, where $n_\mathrm{i}$ is the ion number density.
This effect is therefore only significant for the large radii of clusters \citep{Rudd2009}, whereas the deficit we observe is in group-mass halos.

Cosmic rays \citep[e.g.,][]{Ruszkowski2023} and magnetic fields \citep[e.g.,][]{Parrish2012} have also been proposed as additional sources of non-thermal pressure support.
Recently, \cite{Quataert2025} highlighted that beyond the virial radius of group-mass halos, the pressure from cosmic rays is on the order of the thermal pressure, and cosmic rays could therefore have a significant impact on the thermodynamics of the gas. 
Our results motivate further investigation of these processes in cosmological simulations.

\subsection{Implications for related feedback probes}
\label{sec:disc_implications}

Here we revisit previous constraints on the thermodynamics of groups in light of the thermal pressure deficit found in this work: tSZ stacking measurements, the tSZ power spectrum, and X-ray observations.

Recent measurements of the stacked tSZ effect using the ACT+Planck {\CY} maps have pointed to dramatic pressure deficits ($\gtrsim10\times$) relative to state-of-the-art cosmological hydrodynamical simulations at group masses \citep{Das2025,Das2026Erratum,Liu2025}.
Because these studies consider hydrodynamical simulations that overpredict the kSZ effect \citep[e.g., TNG300 and SIMBA;][]{Hadzhiyska2024photoz,RiedGuachalla2025,Bigwood2025allthesims}, some of the reported discrepancies likely stem from the gas density.
However, our results also suggest that the {\CY} maps adopted by these studies may harbor residual dust contamination, and thus underestimate the tSZ signal;
lastly, we caution that simulation comparisons based on stellar mass estimates are subject to significant systematic uncertainties \citep{McCarthy2025,Siegel2025flamingo, Bigwood2025allthesims}.

Earlier measurements using Planck tSZ data found that the Sloan Digital
Sky Survey ``locally brightest galaxies'' sample is well described by the universal pressure profile (UPP) of \cite{Arnaud2010}, with self-similar mass scaling \citep{Planck2013tsz,Greco2015};
the UPP is a generalized NFW profile fit to low-redshift X-ray data and hydrodynamical simulations.
After accounting for two-halo contributions, which are significant at $M_{500}\lesssim10^{13.5}$~{\Msol} \citep{Vikram2017}, later studies identified a break in the $Y_{500}$--$M_{500}$ relation below $10^{14}$~{\Msol}, and found the strong feedback (AGN~$8.5$) simulations of \cite{LeBrun2015} matched the data well \citep{Lim2018,Hill2018}.
These results are qualitatively consistent with our conclusions; however, we forgo the conversion to $Y_{500}$ to avoid extrapolating below the beam scale and assuming a functional form for the pressure profile, which can introduce significant uncertainties \citep{LeBrun2015}. 

The tSZ power spectrum offers a complementary view of the ICM: it is acutely sensitive to the gas content and thermodynamics of groups and clusters, as well as the cosmological model \citep{Shaw2010,McCarthy2014}.
Relative to the large-scale measurements of Planck \citep{Planck2014tszpower,Bolliet2018}, ACT \citep{Louis2025} and SPT \citep{Raghunathan2026} find a suppression of the tSZ power spectrum at small scales ($\ell \gtrsim 1000$).
The recent joint analysis of Planck, ACT, and SPT data by \cite{Efstathiou2025}, which mitigates CIB and radio contamination by considering the temperature power spectra, rather than {\CY} maps, also points to significant suppression at small scales.
Even the strongest feedback FLAMINGO simulation dramatically overpredicts the observed tSZ power spectrum at $\ell \gtrsim 1000$, apparently requiring both strong feedback and a low $S_8$ cosmology to reproduce the data \citep{Raghunathan2026}.
Our results suggest that some of the suppression may reflect a lack of non-thermal pressure support in state-of-the-art hydrodynamical simulations.

X-ray studies also suggest that groups are strongly shaped by non-gravitational physics. 
For nearby groups, X-ray observations find that the entropy of the gas ($K \propto T_\mathrm{e} n_\mathrm{e}^{-2/3}$) is higher than expected in the gravity-only limit, particularly at smaller radii \citep[$<R_{2500}$;][]{Johnson2009,Sun2009};
drawing conclusions on the level of non-thermal pressure support requires careful accounting of both the gas content and total mass of the systems.
Connecting these observations with the tSZ results is difficult, because X-ray studies are typically biased toward nearby, X-ray-bright systems \citep{Rasmussen2006,Andreon2016,OSullivan2017,Lovisari2021review}.
Indeed, for the SDSS MaxBCG catalog, \cite{Planck2013tsz} found that the $Y$--$M$ relation for the X-ray-bright subsample was systematically higher than for the parent sample. 
Studying the same sample of groups with X-rays and the tSZ effect will be key to building a more complete picture.

\section{Conclusions}
\label{sec:conclusions}

The tSZ effect is a powerful probe of baryonic feedback and non-thermal pressure sources; however, previous studies have yet to converge on a clear picture.
Both the measurement and the theory prediction of the tSZ effect are subject to large systematic uncertainties, which we address in this work: the tSZ signal is contaminated by astrophysical foregrounds, and simulation predictions are acutely sensitive to the halo mass and satellite fraction of the galaxy sample. 
We present new measurements of the tSZ effect around DESI LRGs, spanning $10^{13}<M_{500}<10^{14}$~{\Msol} at $0.4 < z < 1$. 
We benchmark these measurements against the suite of $1$~Gpc$^{3}$ FLAMINGO simulations \citep{Schaye2023}, using joint GGL measurements to ensure a like-with-like comparison.

The main results of this study are as follows:
\begin{enumerate}
    \item Using galaxy-informed SED models of dust and radio contamination, we robustly isolate the tSZ signal.
    We find that dust contamination is significant across the entire mass range, falling from $60\%$ of the $150$~GHz signal within $1.6'$ apertures at $M_{500}=10^{13}$~{\Msol} to $20\%$ at $M_{500}=10^{14}$~{\Msol}.
    Although the radio component is subdominant, its inclusion significantly improves the goodness-of-fit.
    Our SED modeling yields systematically higher {\CY} than the ACT+Planck CIB$+d\beta$ deprojected maps \citep{Coulton2024}.

    \item Forward modeling the tSZ effect from GGL-calibrated selections of simulated galaxies significantly reduces systematic uncertainties in the interpretation of the tSZ measurements.
    Forward modeling accounts for the effects of miscentering and the two-halo term, and marginalizes over mass and redshift.
    We demonstrate that for small apertures ($\lesssim3'$), the GGL-calibrated simulation predictions are robust to large variations in the galaxy selection procedure.
    \item The fiducial ({\fidsim}) FLAMINGO simulation overpredicts {\CY} across the full mass ($10^{13} < M_{500} < 10^{14}$~{\Msol}) and redshift ($0.4 < z < 1$) range of the photometric LRG sample;
    the discrepancy is largest at small apertures: $\lesssim3'$, which corresponds to $4\,R_{500}$ at $z=0.7$.
    The strongest feedback ({\eightsigma}) FLAMINGO simulation provides a better match to the data but still overpredicts {\CY} by $\approx 2 \times$ at $1.6'$.
    \item Because the strongest feedback ({\eightsigma}) FLAMINGO simulation successfully reproduces the gas density around DESI LRGs \citep{Siegel2025flamingo,Bigwood2025allthesims}, as traced by DESI+ACT kSZ measurements \citep{Hadzhiyska2024photoz,RiedGuachalla2025}, the discrepancy with the tSZ effect points to the simulated gas temperature being too high.
    At these scales ($\gtrsim R_{500}$), the simulated gas is in approximate hydrostatic equilibrium. 
    Maintaining equilibrium, while lowering the gas temperature at fixed density, thus requires significant non-thermal pressure support, beyond the kinetic contribution already realized in the simulations.
\end{enumerate}

\paragraph{\textbf{Outlook.}} In this work, we have confronted the key uncertainties in measuring and interpreting the tSZ effect.
The Simons Observatory, which covers $30$--$280$~GHz with six channels, will further improve our ability to isolate the tSZ signal from astrophysical contaminants \citep{Ade2019}. 
Forward modeling the dust and radio emission from simulated halos will also be informative for developing improved separation methods and investigating residual systematics.

The observed thermal pressure deficit motivates further investigation of non-thermal pressure sources in cosmological simulations---particularly cosmic rays \citep[e.g.,][]{Ruszkowski2023} and magnetic fields \citep[e.g.,][]{Parrish2012}---and of how much groups deviate from hydrostatic equilibrium.
Characterizing the intrinsic scatter in the hot gas properties---as a function of mass, redshift, and scale---and its correlations with galaxy and ICM properties will be key to understanding the physical drivers of the deficit.
Such investigations are also motivated by other probes.
Recent kSZ effect measurements from SDSS/DESI+ACT and eROSITA X-ray gas mass fractions require stronger baryon feedback at group masses than realized in most state-of-the-art hydrodynamical simulations \citep{McCarthy2025,Siegel2025flamingo,Bigwood2025allthesims}, but current strong feedback variants appear to deviate from X-ray cluster scaling relations \citep{Braspenning2024,Eckert2026}.
The growing landscape of observational probes suggests that a re-examination of the physics governing the gas in groups is needed. 

\acknowledgments
JS acknowledges support by the National Science Foundation Graduate Research Fellowship Program under Grant DGE-2039656. 
Any opinions, findings, and conclusions or recommendations expressed in this material are those of the author(s) and do not necessarily reflect the views of the National Science Foundation.
JS acknowledges helpful conversations with Tianyi Yang and George Efstathiou that contributed to the quality of this work.
We thank Zachary Atkins for helpful guidance in using the ACT data. 

\appendix

\section{Simulation variants}
\label{appendix:simulation_variants}

To compare the FLAMINGO simulations with the tSZ measurements, we select samples of simulated galaxies that best fit the observed GGL profiles (Section~\ref{sec:forward_model}).
Our fiducial analysis draws centrals and satellites from a log-normal stellar mass distribution with $\sigma=0.2$~dex and omits simulated halos with $M_{500}>10^{14.3}$~{\Msol}.
The width of the log-normal distribution is motivated by the width of the stellar mass bins and systematic uncertainties in the stellar mass estimates \citep{Siegel2025flamingo},
and the masking of massive clusters mirrors our removal of ACT-detected clusters in the measurements (also see Appendix~\ref{appendix:tsz_clusters}).
The center of the stellar mass distribution is determined by the GGL fit.

\begin{figure*}[t!]
\includegraphics[width=\textwidth]{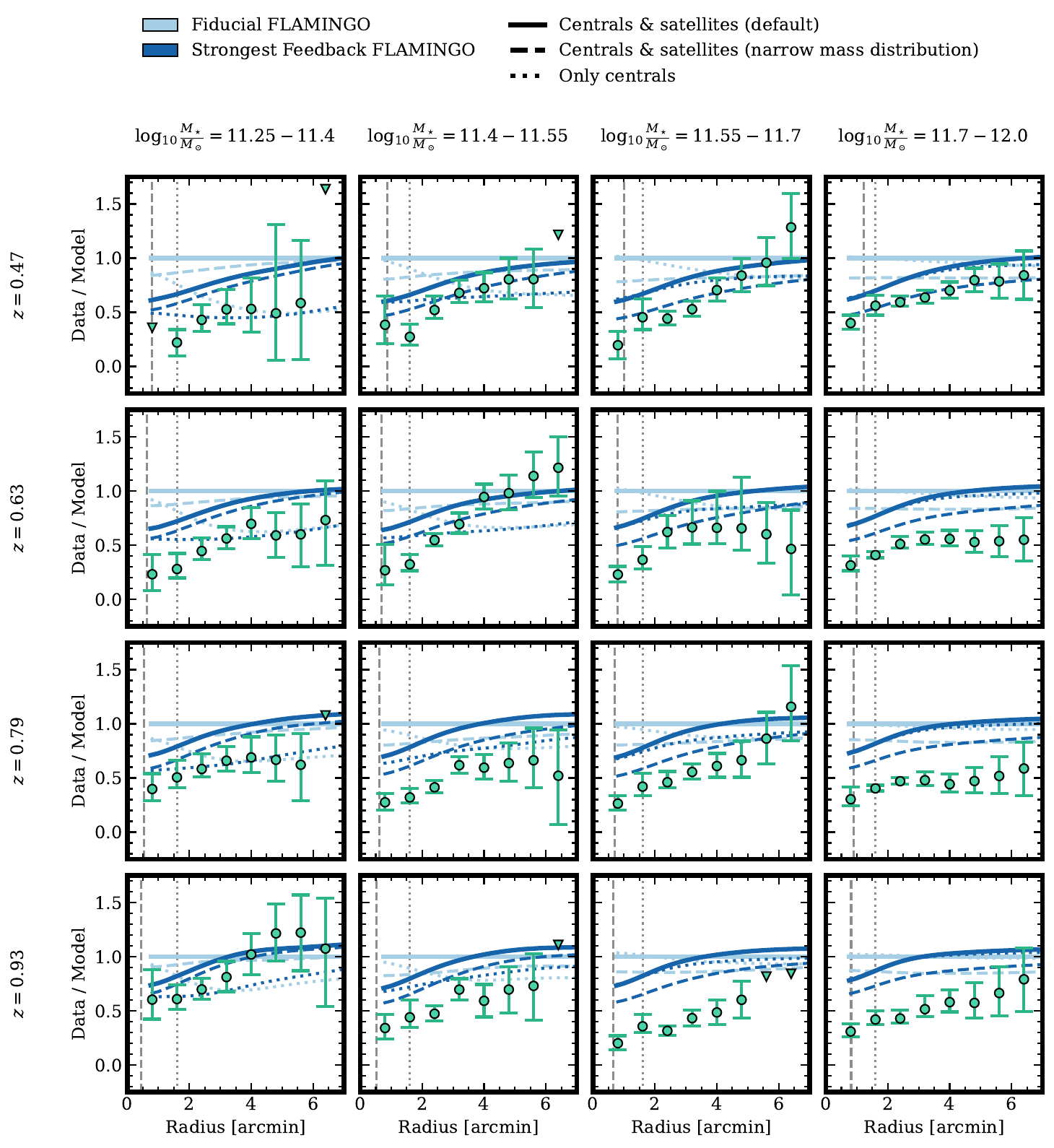}
\caption{
The {\CY} profiles (green) relative to the prediction of the fiducial ({\fidsim}; solid light blue) FLAMINGO simulation;
the strongest feedback simulation ({\eightsigma}) is shown in dark blue.
The dotted and dashed lines present simulation predictions without satellite galaxies and with narrow stellar mass distributions ($\sigma=0.05$~dex), respectively; the GGL calibration is repeated for each variant. 
Each row corresponds to a different redshift bin, and each column presents a different stellar mass bin.
Where the tSZ constraint is consistent with zero, we show the $2\sigma$ upper limit as a triangle.
The dashed vertical lines present $R_{500}$ for each sample, and the vertical dotted lines mark $1.6'$: the beam FWHM and the fiducial scale of our primary analysis.
}
\label{fig:grid_comp}
\end{figure*}

To explore the maximal range of simulation predictions, we consider two extreme (and physically implausible) variants of the fiducial selection function: (i) centrals-only samples and (ii) ultra-narrow stellar mass distributions ($\sigma=0.05$~dex); the GGL calibration is repeated for each variant.
Figure~\ref{fig:grid_comp} compares the simulation predictions against the tSZ measurements for each mass and redshift bin.
By including fewer massive halos, the variants lower the tSZ signal relative to our fiducial selection.
Because such extreme cases are implausible for the DESI LRGs, these predictions are informative lower bounds on the tSZ signal.
Even these modeling choices do not fully resolve the deficit; see Section~\ref{sec:disc_robustness} for an extended discussion.

\section{Astrophysical Contaminants}
\label{appendix:contaminants}

The fiducial tSZ constraints are derived from a joint tSZ, dust, and radio SED model, assuming $T_\mathrm{dust}=24$~K and $\beta=1.7$.
In this appendix, we demonstrate that our conclusions are robust to variations of the SED model---a more flexible dust component and the omission of the radio component---and to the exclusion of radio galaxies from the sample.

\begin{figure*}[t!]

\gridline{ \fig{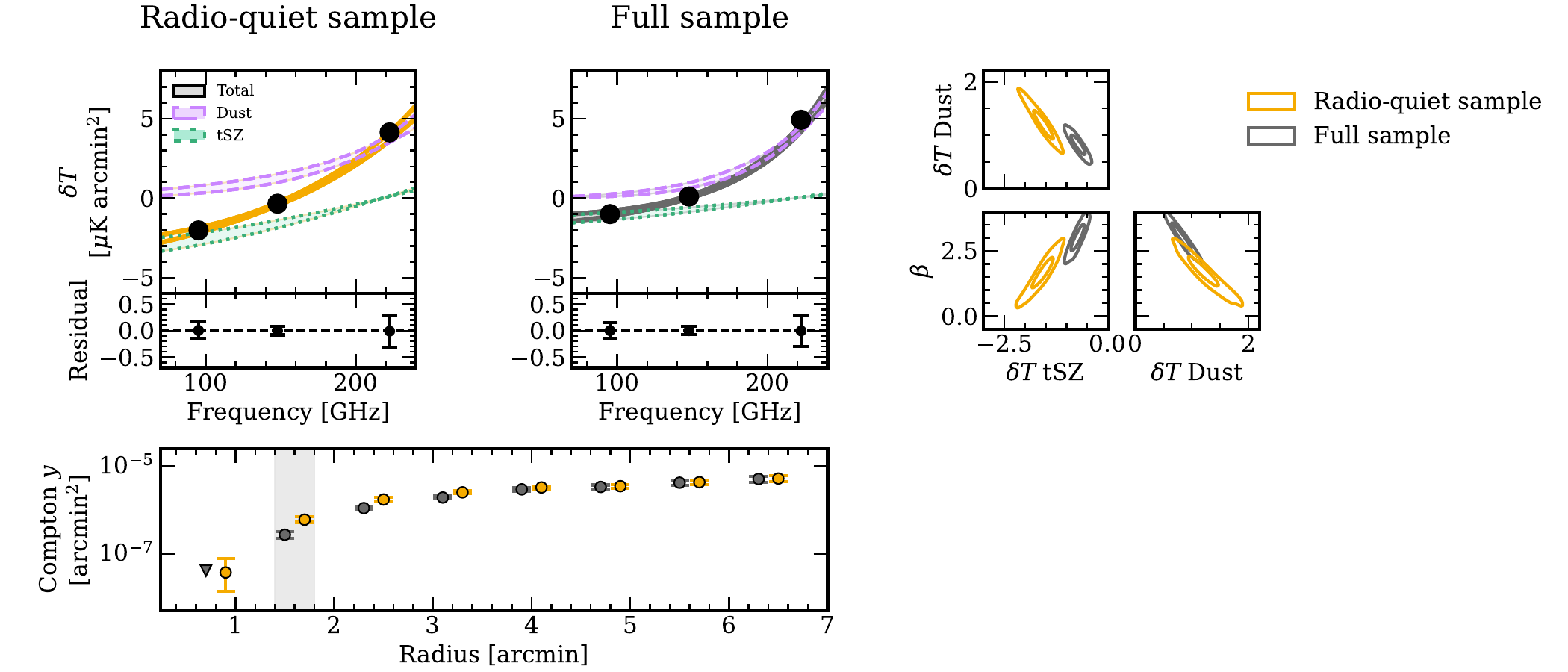}{\textwidth}{} }

\caption{
Comparison of the SED models for the radio-quiet and full LRG samples, assuming a joint tSZ and dust model; the full LRG sample favors extreme $\beta$ of $\gtrsim2.5$, likely reflecting the need for a radio component.
Left and middle columns: stacked temperature fluctuations as a function of frequency for the radio-quiet LRG sample (left; yellow) and the full sample with no radio cut (middle; gray), shown for the $1.6'$ aperture.
Both samples are at $z=0.54$--$0.71$ with $\log_{10} M_\star / M_\odot=11.25$--$12.0$.
For both samples, $\beta$ is a free parameter and $T_\mathrm{dust}=24$~K.
Beneath the SEDs, we present the inferred {\CY} profiles;
points are offset horizontally for clarity, and the gray band marks the $1.6'$ aperture (beam FWHM). 
Right column: posteriors of the SED fits.
}
\label{fig:breakdown_without_radiocomp}
\end{figure*}

\begin{figure*}[t!]

\gridline{ \fig{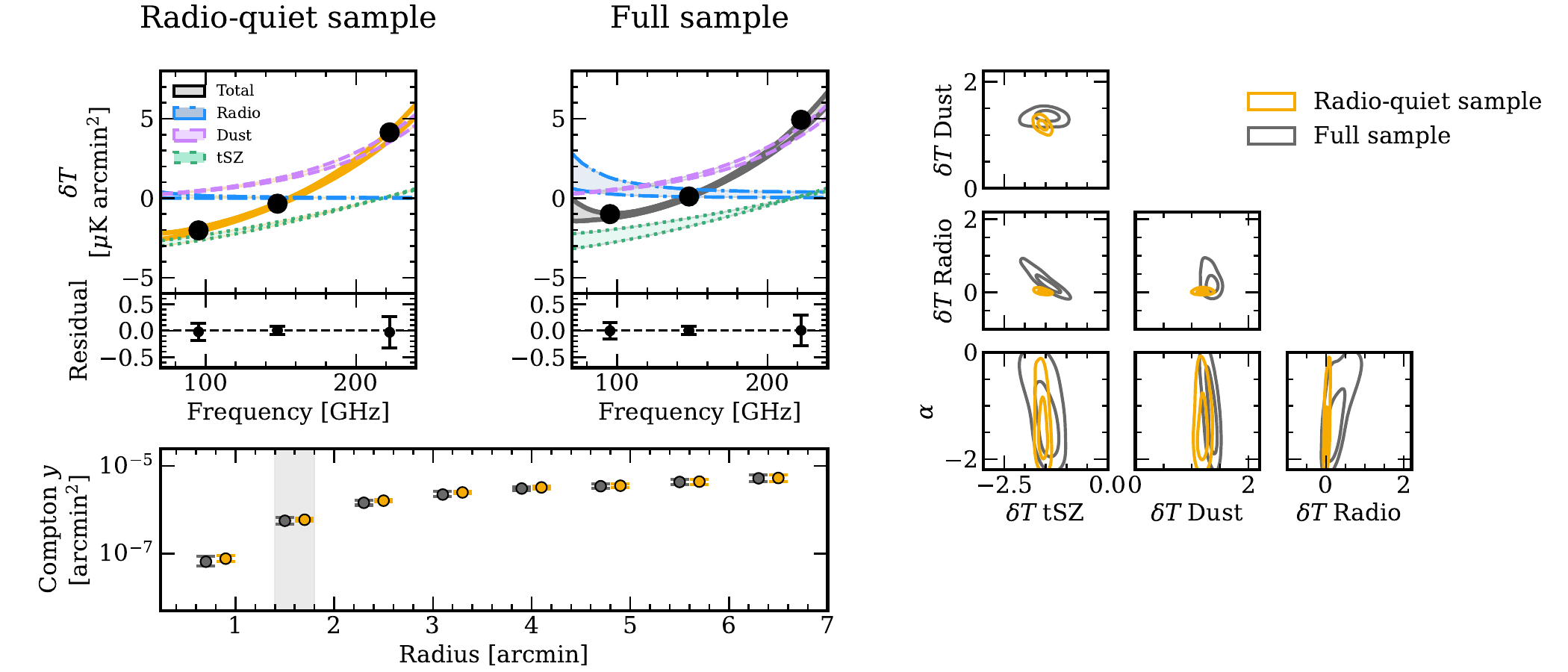}{\textwidth}{} }

\caption{
Comparison of the SED models for the radio-quiet and full LRG samples, assuming the fiducial joint tSZ, dust, and radio model with $\beta=1.7$ and $T_\mathrm{dust}=24$~K. 
Left and middle columns: stacked temperature fluctuations as a function of frequency for the radio-quiet LRG sample (left; yellow) and the full sample with no radio cut (middle; gray), shown for the $1.6'$ aperture.
Both samples are at $z=0.54$--$0.71$ with $\log_{10} M_\star / M_\odot=11.25$--$12.0$.
Beneath the SEDs, we present the inferred {\CY} profiles;
points are offset horizontally for clarity, and the gray band marks the $1.6'$ aperture (beam FWHM). 
Right column: posteriors of the SED fits.
}
\label{fig:breakdown_with_radiocomp}
\end{figure*}

Our fiducial dust parameters ($T_\mathrm{dust}=24$~K and $\beta=1.7$) are informed by the subsample of radio-quiet LRGs, where the tSZ effect and dust emission dominate the signal.
The joint tSZ and dust model infers $\beta = 1.6 \pm 0.7$, $1.6 \pm 0.6$, $1.8 \pm 0.5$, and $1.8 \pm 0.6$ for the four redshift bins, respectively.
The posteriors of the model are shown in Figure~\ref{fig:breakdown_without_radiocomp} for a representative sample of radio-quiet galaxies.
In Section~\ref{sec:sed_modeling}, we found that the full sample is poorly described by the tSZ and dust model with $\beta = 1.7$, motivating the inclusion of a radio component.
If $\beta$ is freed, the tSZ and dust model can fit the data but only for $\beta \gtrsim 2.5$ (Figure~\ref{fig:breakdown_without_radiocomp}).
Observational \citep[e.g.,][]{Addison2013,Planck2014CIB,Yang2026} and theoretical \citep{Draine1984} studies typically favor $\beta<2$;
however, higher values could arise from amorphous grains \citep{Meny2007}, and a multi-temperature dust distribution can mimic a high $\beta$ \citep{Shang2012}. 
Because such high $\beta$ are disfavored for similar galaxy samples \citep[e.g., eBOSS LRGs;][]{Chiang2025} and the sample is known to include radio sources, we adopt the joint tSZ, dust, and radio model with $\beta=1.7$ for our fiducial analysis. 
Figure~\ref{fig:breakdown_with_radiocomp} presents the posteriors of the fiducial model, for the same representative sample of galaxies described above. 

\begin{figure*}[t!]

\gridline{ \fig{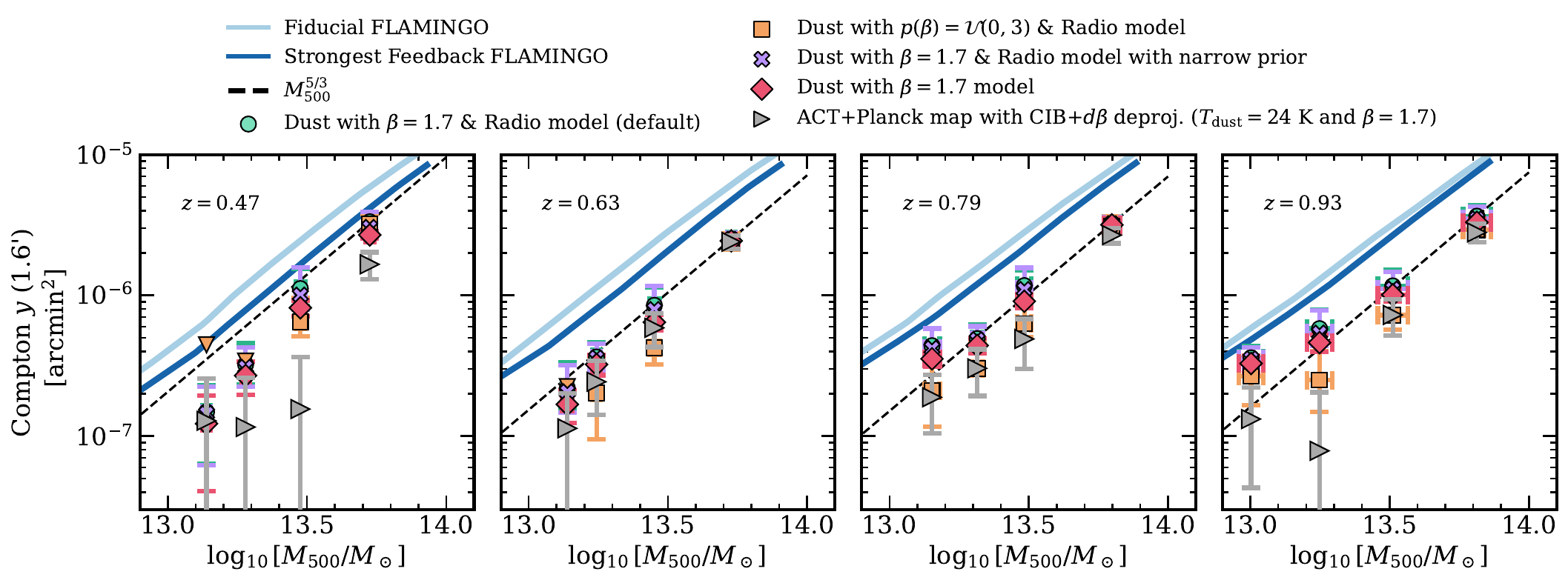}{\textwidth}{} }

\caption{
The {\CY} constraints are robust to large variations in the SED modeling process. 
The stacked {\CY} amplitude at $1.6'$ is shown as a function of mean halo mass for the fiducial SED model (green circles) and three variants: (i) treating $\beta$ as a free parameter (orange squares), (ii) imposing a VLASS-informed prior (purple crosses), and (iii) omitting the radio component (red diamonds).
Where the tSZ constraint is consistent with zero, we show the $2\sigma$ upper limit as a triangle.
For reference, we also include the stacked {\CY} from the CIB$+d\beta$ deprojected map with the same dust parameters as our fiducial SED model ($T_\mathrm{dust}=24$~K and $\beta=1.7$) as rightward-facing gray triangles.
Each column presents a different redshift bin.
Predictions from the fiducial ({\fidsim}) and strongest feedback ({\eightsigma}) FLAMINGO simulations are shown as light and dark blue lines, respectively.
The dashed line presents the $M_{500}^{5/3}$ self-similar scaling with an arbitrary normalization. 
}
\label{fig:m_v_z_comp}
\end{figure*}

The tSZ measurements are robust to large changes in the dust and radio components. 
In Figure~\ref{fig:m_v_z_comp}, we compare the fiducial {\CY} constraints for three variants of the SED model:
(i) treating $\beta$ as a free parameter, (ii) imposing a VLASS-informed prior on the amplitude of the radio component, and (iii) omitting the radio component.
The constraints are consistent at the $1\sigma$ level for every mass and redshift bin.
The fiducial model yields the highest {\CY}.
It is therefore a conservative choice for our simulation comparison, as well as being preferred on astrophysical grounds (Section~\ref{sec:sed_modeling}).

\begin{figure*}[t!]

\gridline{ \fig{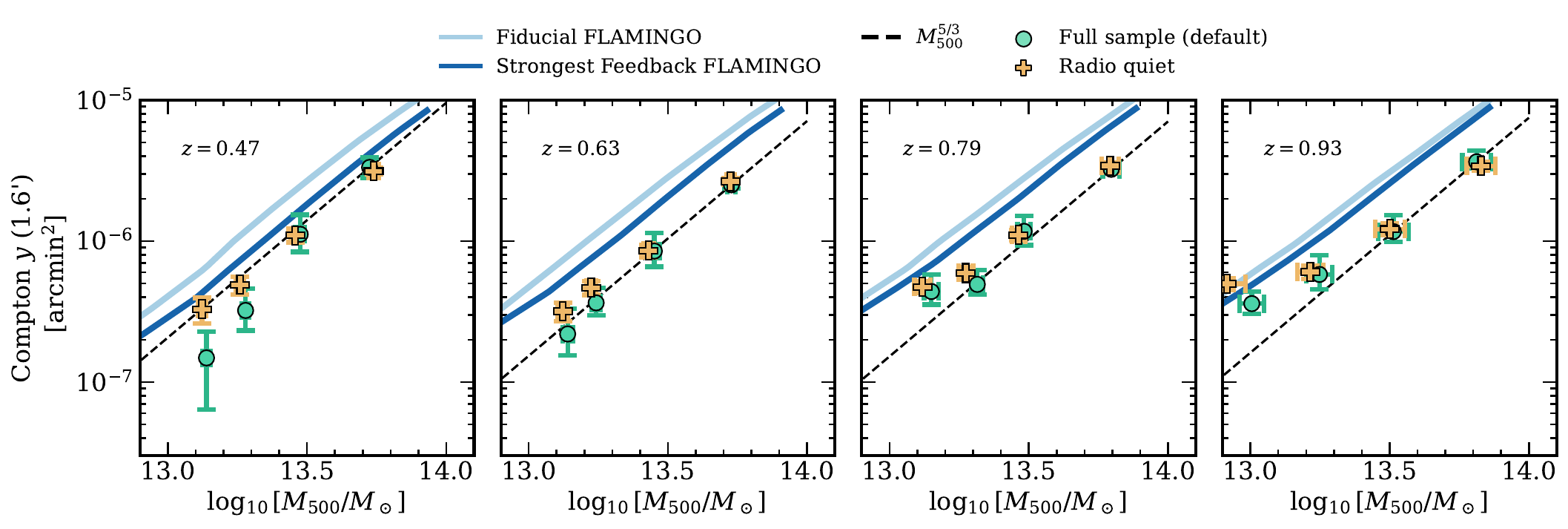}{\textwidth}{} }

\caption{
Comparison of the stacked {\CY} amplitude at $1.6'$ for the full LRG sample (green circles) and the radio-quiet subsample (yellow pluses).
The radio-quiet and full samples are consistent at the $1\sigma$ level; however, the {\CY} values for the full sample are systematically lower, potentially reflecting differences in the gas distribution or residual radio contamination.
The {\CY} constraints are derived from the joint tSZ, dust, and radio model.
Each column presents a different redshift bin.
Predictions from the fiducial ({\fidsim}) and strongest feedback ({\eightsigma}) FLAMINGO simulations are shown as light and dark blue lines, respectively.
The dashed line presents the $M_{500}^{5/3}$ self-similar scaling with an arbitrary normalization. 
}
\label{fig:m_v_z_radio}
\end{figure*}

The {\CY} constraints for the radio-quiet and full samples are consistent at the $1\sigma$ level.
Figure~\ref{fig:breakdown_with_radiocomp} compares the posteriors for a representative sample with and without the radio-quiet criterion.
Although the constraints are statistically consistent, the model yields systematically higher {\CY} for the radio-quiet samples.
For each mass and redshift bin, Figure~\ref{fig:m_v_z_radio} compares the {\CY} constraints for the radio-quiet and full samples as a function of mass;
the GGL measurements are repeated for the radio-quiet subsamples.
At fixed stellar mass, the radio-quiet galaxies reside in slightly less massive halos, qualitatively consistent with previous findings \citep[e.g.,][]{Mandelbaum2009}.
The difference in {\CY} between the radio-quiet and full samples could reflect physical differences in the gas and/or residual radio contamination; also see \cite{Battaglia2026}.
With the available data, the difference is small and does not alter our conclusions.
Revisiting this question with the Simons Observatory is warranted.


\section{ACT tSZ Clusters}
\label{appendix:tsz_clusters}

\begin{figure}[t!]
\centering
\includegraphics[width=0.4\columnwidth]{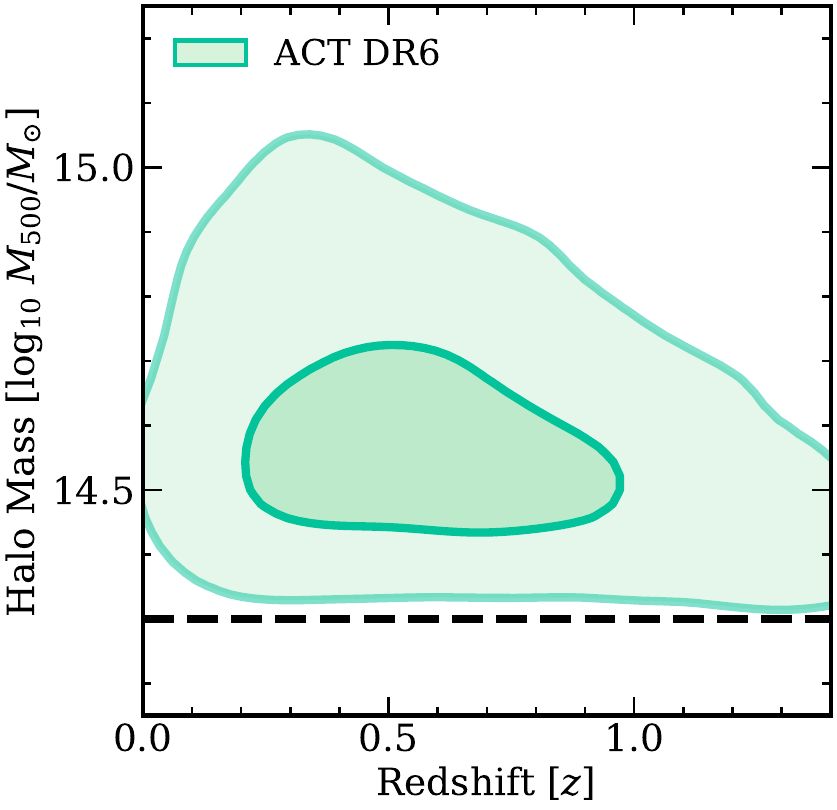}
\caption{
The distribution of ACT DR6 clusters ($S/N>6$) in mass and redshift from \cite{Aguena2026};
the halo masses are estimated from a weak lensing calibrated scaling relation.
The contours represent the $50$th and $95$th percentiles.
The dashed horizontal line presents the criterion applied to the simulated halos: $M_{500}<10^{14.3}$~{\Msol}.
}
\label{fig:ACTclusters}
\end{figure}

We omit all DESI LRGs within $15'$ of ACT-detected clusters from our stacking analysis. 
Following \cite{Liu2025}, this masking suppresses the outsized contribution of clusters to the stacked signal. 
We adopt the ACT~DR6 cluster catalog \citep{Aguena2026} and apply a minimum $S/N$ threshold of $6$.
Figure~\ref{fig:ACTclusters} presents the mass and redshift distribution of ACT clusters;
the halo masses are estimated from a weak lensing calibrated scaling relation between mass, {\CY}, and redshift.
Because the weak lensing calibrations are limited to $z\lesssim0.8$ and the hydrostatic mass bias increases with redshift \citep{Robertson2024,Shin2025}, the masses are likely underestimated at higher redshifts.

To mirror this masking in the simulation forward modeling, we omit all simulated halos with $M_{500}>10^{14.3}$~{\Msol} (dashed horizontal line in Figure~\ref{fig:ACTclusters}).
This criterion is intentionally stricter than the masking applied to the data:
the $10^{14.3}$~{\Msol} threshold lies below the ACT cluster mass distribution, and our mask removes \textit{every} simulated halo above the threshold, whereas the ACT catalog is incomplete \citep[see Figure 8 of][]{Aguena2026}.
The tSZ predictions from the simulation are therefore conservatively low; relaxing the masking criterion would increase the apparent pressure deficit.

\section{Galaxy-Galaxy Lensing}
\label{appendix:ggl}

We measure the GGL profiles of the DESI LRGs in bins of mass and redshift using the HSC~Y3 shear catalog.
The Hyper Suprime-Cam (HSC) Subaru Strategic Program is conducting a 300-night $grizy$ imaging survey with the $8.2$~meter Subaru telescope \citep{Aihara2018}.
We use the three-year release\footnote{\url{https://hsc-release.mtk.nao.ac.jp/doc/}} \citep{Li2022}.
HSC is adopted because it includes sufficiently high-redshift tomographic bins to measure the GGL signal across the entire LRG redshift distribution.
HSC was previously shown to be consistent with DES~Y3 and KiDS~1000, after correcting the mean redshifts of the HSC tomographic bins \citep{Heydenreich2025}; also see \cite{Amon2023}.
We include the tomographic bin correction following \cite{Li2023} and \cite{Heydenreich2025}.

The GGL signal is measured following the methodology of \citet{Lange2024} and \citet{Heydenreich2025}, briefly outlined below.
The GGL measurements are presented in Figure~\ref{fig:grid_ggl}, alongside the best-fitting FLAMINGO GGL profiles.

The excess surface density measurements are corrected for known systematics, including shear calibration and lens magnification bias; see Appendix~A of \cite{Siegel2025flamingo} for an extended discussion.
Because randoms catalogs are unavailable for the photometric LRG sample, we forgo randoms subtraction and boost factor corrections.
For the spectroscopic LRG sample, we find the randoms signal is consistent with zero, and the boost correction only rises above the $2\%$ level for small separations ($<0.5$~comoving~{\Mpch}) in the highest redshift bin;
shifting the GGL signal by $2\%$ corresponds to $0.01$~dex in mean halo mass. 
For the first three redshift bins, we consider $>0.2$~comoving~{\Mpch}, and for the last bin, we consider $>0.5$~comoving~{\Mpch}.
The covariance matrices are measured via two-dimensional leave-one-out jackknife sampling.
The jackknife patches are approximately $20$ comoving~{\Mpch} at the mean redshift of the LRGs.
We therefore restrict our GGL fitting to $<20$~comoving~{\Mpch}.

\begin{figure}[t!]
\centering
\includegraphics[width=0.5\columnwidth]{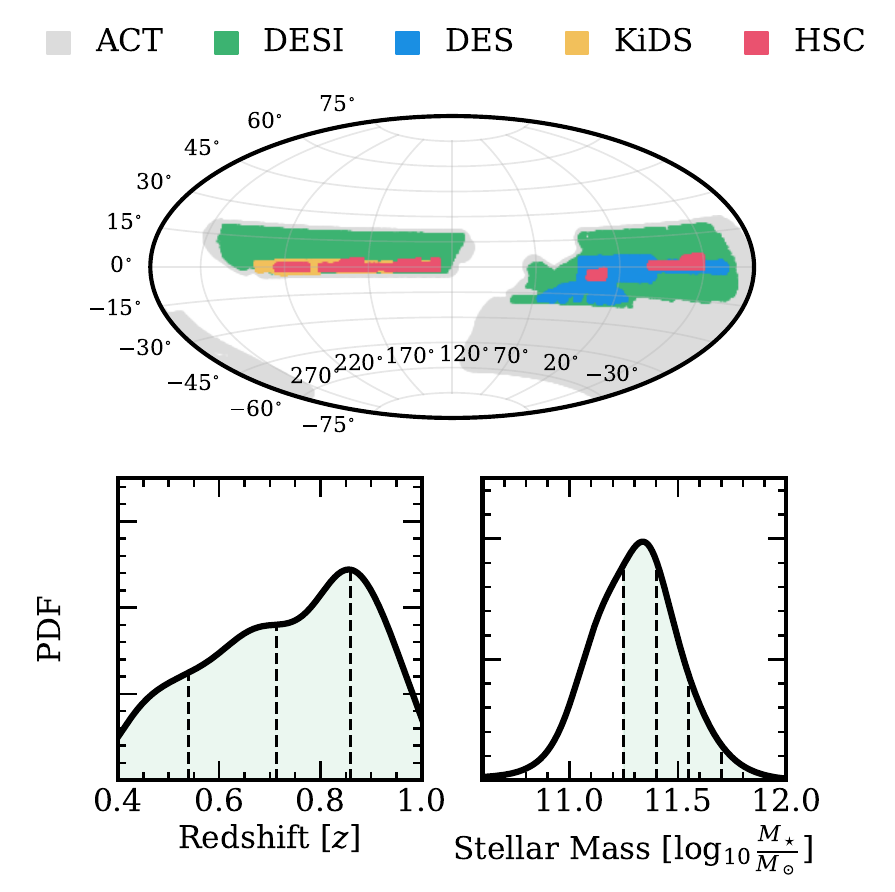}
\caption{
\textit{Top:} the footprints of the DESI photometric LRG sample (green) and ACT DR6 (gray) alongside the HSC (red) lensing survey; we also include the DES (blue) and KiDS (yellow) lensing surveys for context.
The DESI and lensing survey footprints are limited to their regions of overlap with ACT.
\textit{Bottom:} The redshift distribution (left) and stellar mass distributions (right) of the photometric DESI LRG sample. 
The vertical dashed lines represent the four redshift bin edges \citep[following][]{Zhou2023}
 and the mass bin edges; the mass bins are limited to $\log_{10}M_\star / M_\odot>11.25$.
}
\label{fig:survey_map}
\end{figure}

\begin{figure*}[t!]
\includegraphics[width=\textwidth]{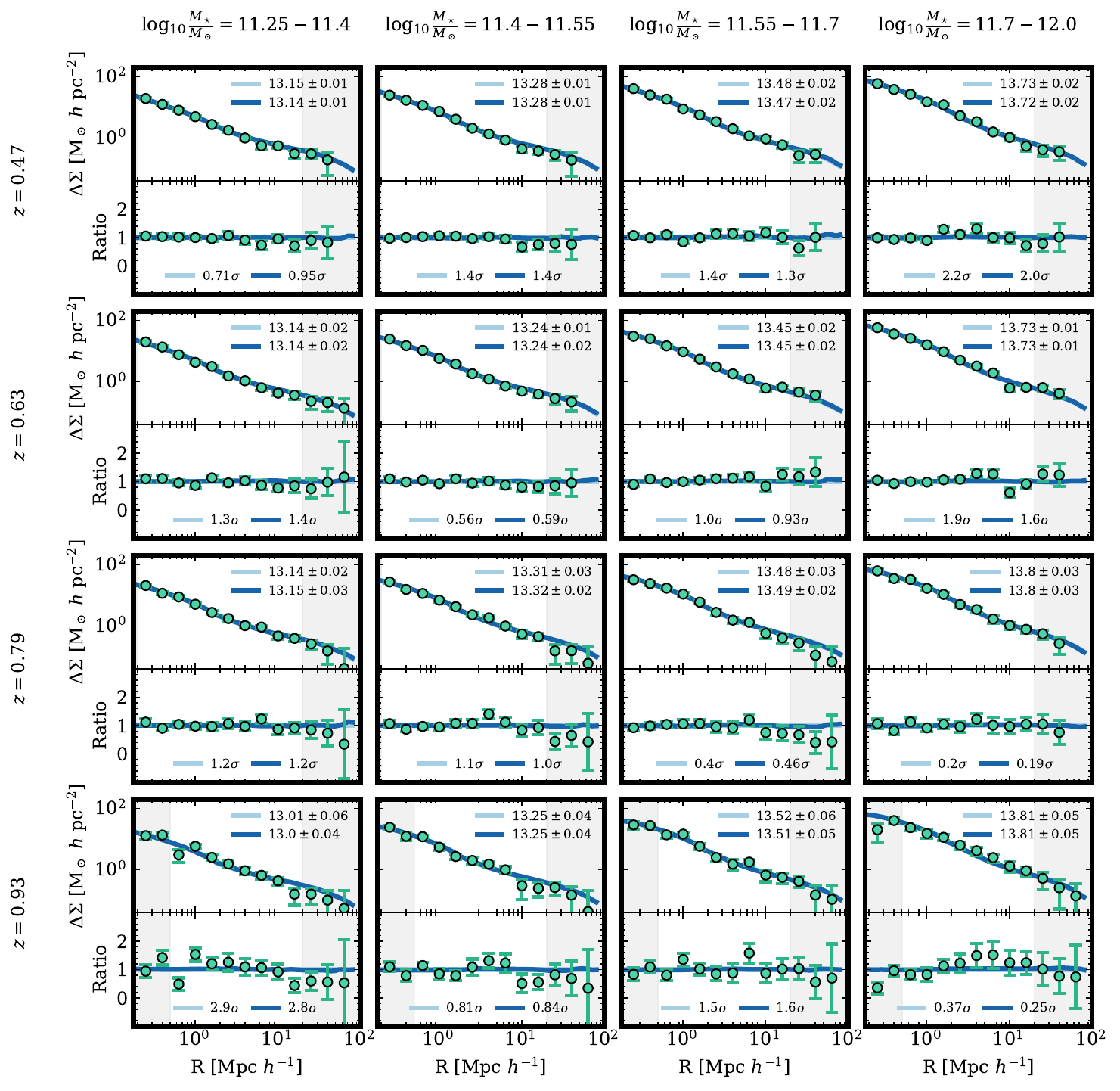}
\caption{
The excess surface density measurements (green) alongside the best-fitting FLAMINGO GGL profiles for the fiducial ({\fidsim}; light blue) and strongest feedback ({\eightsigma}; dark blue) simulations.
The lower panels show the ratio relative to the fiducial FLAMINGO simulation and report the number of standard deviations by which each simulation deviates from the observations.
Each row corresponds to a different redshift bin, and each column presents a different stellar mass bin.
}
\label{fig:grid_ggl}
\end{figure*}

\bibliography{paper}%

\end{document}